\documentclass[12pt,a4paper]{article}
\usepackage{jheplikemod,tocloft,amsmath,textcomp,mathrsfs,mathtools,graphicx,bbm,epigraph,tikz}
\usepackage[bb=libus]{mathalpha}
\usepackage[T1]{fontenc}
\usepackage{lmodern}
\hypersetup{pdftitle={Complete Prescriptivity},pdfcreator={},linkcolor=[rgb]{0.15,0.35,0.75},colorlinks=true,citecolor=[rgb]{0.675,0,0.2},urlcolor=[rgb]{0.15,0.35,0.65}}
\graphicspath{{figures/}}
\let\olditemize\itemize\renewcommand{\itemize}{\vspace{-9pt}\olditemize\setlength{\itemsep}{0.5pt}\setlength{\parskip}{0pt}\setlength{\parsep}{-2pt}}
\let\oldenumerate\enumerate\renewcommand{\enumerate}{\vspace{-4pt}\oldenumerate\setlength{\itemsep}{1pt}\setlength{\parskip}{0pt}\setlength{\parsep}{0pt}}

\makeatletter\renewcommand\section{\addtocontents{toc}{\protect~\\[-32pt]}\@startsection {section}{1}{\z@}{0.5ex \@plus .2ex \@minus 1.0ex}{1.ex \@plus.0ex\@minus 1.0ex}{\normalfont\large\bfseries}}
\renewcommand\subsection{\addtocontents{toc}{\protect\addvspace{-1.5\p@}}\@startsection {subsection}{1}{\z@}{1.5ex \@plus .2ex \@minus 2.0ex}{0.5ex \@plus.1ex\@minus 2.5ex}{\normalfont\bfseries}}
\renewcommand\subsubsection{\addtocontents{toc}{\protect\addvspace{-2.5\p@}}\@startsection {subsubsection}{1}{\z@}{1.5ex \@plus .2ex \@minus 2.0ex}{1.ex \@plus.1ex\@minus 2.5ex}{\normalfont\bfseries}}
\definecolor{rindou1}{rgb}{0.4431,0.2862,0.7960}
\definecolor{rindou2}{rgb}{0.0078,0.1215,0.4392}
\definecolor{lapis}{rgb}{0.0.0470,0.2941,0.5568}
\definecolor{emerald}{rgb}{0.31, 0.78, 0.47}
\definecolor{pinegreen}{rgb}{0.0, 0.47, 0.44}
\definecolor{jade}{rgb}{0.0, 0.66, 0.42}
\definecolor{teal}{rgb}{0.0, 0.5, 0.5}
\definecolor{hblue}{rgb}{0,0,0.575}
\definecolor{hred}{rgb}{0.575,0.0,0.225}
\definecolor{hgreen}{rgb}{0.0,0.4,0.2}
\definecolor{hteal}{rgb}{0.0,0.445,0.6451}
\definecolor{varcolor}{rgb}{0.1,0.55,0.25}
\definecolor{functioncolor}{rgb}{0.1,0.35,0.75}

\newcounter{topologyIndex}
\newcounter{topology}
\newcommand{\eq}[1]{\vspace{-0.5pt}\begin{equation}#1\vspace{-0.5pt}\end{equation}}

\newcommand{\fwbox}[2]{\text{\makebox[#1][c]{$\hspace{-150pt}\displaystyle#2\hspace{-150pt}$}}}
\newcommand{\fwboxL}[2]{\text{\makebox[#1][l]{$#2$}}}
\newcommand{\fwboxR}[2]{\text{\makebox[#1][r]{$#2$}}}
\newcommand{\equivR}{\fwbox{14.5pt}{\hspace{-0pt}\fwboxR{0pt}{\raisebox{0.47pt}{\hspace{1.25pt}:\hspace{-4pt}}}=\fwboxL{0pt}{}}}
\newcommand{\equivL}{\fwbox{14.5pt}{\fwboxR{0pt}{}=\fwboxL{0pt}{\raisebox{0.47pt}{\hspace{-4pt}:\hspace{1.25pt}}}}}
\newcommand{\fig}[3]{\raisebox{#1}{$\begin{array}{@{}c@{}}\includegraphics[scale=#2]{#3}\end{array}$}}
\newcommand{\Bigger}[1]{\raisebox{-2.25pt}{\scalebox{1.75}{$#1$}}}
\newcommand{\bigger}[1]{\raisebox{-0.95pt}{\scalebox{1.25}{$#1$}}}

\renewcommand{\phi}{\varphi}

\renewcommand{\hat}{\widehat}

\newcommand{\x}[2]{{\color{black}(}\hspace{-0.85pt}{\color{black}#1}\hspace{-0.25pt}{\color{black}|}\hspace{-0.25pt}{\color{black}#2}\hspace{-0.85pt}{\color{black})}}
\newcommand{\dbar}{\fwboxL{7.2pt}{\raisebox{4.5pt}{\fwboxL{0pt}{\scalebox{1.5}[0.75]{\hspace{1.25pt}\text{-}}}}d}}
\renewcommand{\r}[1]{{\color{hred}#1}}
\renewcommand{\b}[1]{{\color{hblue}#1}}

\renewcommand{\t}[1]{{\color{hteal}#1}}

\renewcommand{\brace}[1]{\hspace{-2pt}{\texttt{[}}#1{\texttt{]}}\hspace{-1pt}}
\newcommand{\pattern}{{\color{varcolor}\rule[-1.05pt]{7.5pt}{.65pt}}}

\renewcommand{\inf}{\b{X}}
\newcommand{\indices}[2]{{\hspace{-0.0pt}}^{\smash{#1}}_{\phantom{\smash{#1}}\smash{#2}}}
\newcommand{\fun}[1]{\text{\texttt{\textbf{\texttt{\color{functioncolor}#1}}}}}
\newcommand{\uscore}{\rule[-1.05pt]{7.5pt}{.75pt}}
\newcommand{\var}[1]{{{\color{varcolor}{\sl#1}}}}
\newcommand{\built}[1]{{\color{black}\textbf{\texttt{#1}}}}

\makeindex
\title{\texorpdfstring{\mbox{{\huge \emph{Complete} Prescriptivity:}{\LARGE}}\\[-5pt]\mbox{{\huge Matching \emph{Every} Cut of sYM}}}{Complete Prescriptivity}}
\author[\hspace{-0pt}\dagger,\ast]{\vspace{-24pt} Jacob~L.~Bourjaily}\emailAdd{jbourj@ias.edu}

\affiliation[\dagger]{School of Natural Sciences, Institute for Advanced Study, Princeton, NJ, 08540, USA} 
\affiliation[\ast]{Institute for Gravitation and the Cosmos, Department of Physics,\\Pennsylvania State University, University Park, PA 16802, USA}

\abstract{%
Recently, a basis of individually pure, prescriptive master integrals was described \cite{Bourjaily:2026icc} that spans all maximal-weight, planar contributions to amplitudes in four-dimensional theories at two loops. We describe the representation of amplitudes in planar, maximally supersymmetric Yang-Mills theory in this basis, and contrast this representation with ones that manifest the dual-conformality of the theory. In this new representation, every cut supported by the theory is individually matched, and ratio functions are \emph{manifestly finite}: expressible in terms of integrals that are individually free of any region of infrared divergence.\\[-5pt]

In the ancillary files attached to this work's submission to the \texttt{arXiv}, we have implemented this new representation together with a reformulation of existing representations at one and two loops. 
}

\preprint{}

\begin{document}

\maketitle\thispagestyle{empty}
\pagenumbering{roman}\clearpage

%================================================================================================================
%================================================================================================================
\addtocontents{toc}{\protect~\\[-24.5pt]}
\setcounter{section}{0}

\newpage
\pagenumbering{arabic}
\vspace{0pt}%
\section{Introduction and Overview}\label{sec_introduction}\vspace{-4pt}

In recent decades, many profound discoveries about the structure and form perturbative quantum field theory have emerged from the study and explicit computation of scatting amplitudes. Such discoveries include the discovery of unanticipated symmetries such as dual-conformal invariance \cite{Drummond:2006rz,Alday:2007hr} and the Yangian \cite{Drummond:2009fd,Drummond:2010uq,Drummond:2010qh,Beisert:2010gn}, the correspondence between the cuts of perturbative amplitudes and Grassmannian geometry \cite{ArkaniHamed:2009dn,Kaplan:2009mh,ArkaniHamed:2009dg,ArkaniHamed:2009sx,ArkaniHamed:2009vw,ArkaniHamed:2010gg,ArkaniHamed:book} and the geometric formulation of amplitudes in the form of the amplituhedron \cite{Arkani-Hamed:2013jha,Arkani-Hamed:2013kca,Damgaard:2019ztj}, surprising connections between gauge-theory and gravity \cite{KLT,Bern:2008qj,Carrasco:2011mn,Carrasco:2012ca,Bjerrum-Bohr:2013iza,Monteiro:2013rya,He:2015wgf,Mogull:2015adi,Johansson:2017bfl,Bern:2017yxu,Bern:2017ucb,Edison:2022jln,Porkert:2022efy}, and a great deal more.

Almost all of these examples owe their origins to concrete `theoretical data' in the form of concrete, analytic, machine-readable (and manipulable) expressions for explicit targets of investigation, often in especially simple quantum field theories \cite{Arkani-Hamed:2008owk}---with implications well beyond the narrow corners of their initial manifestations.\\[-6pt]

In this work, we make use of the complete basis of planar, two-loop master integrals of maximal-weight \cite{Bourjaily:2026icc} to provide a new representation of all amplitudes in maximally supersymmetric Yang-Mills theory (`sYM'); we show that this representation allows \emph{every} cut supported by field theory to be matched individually, and results in a \emph{manifestly finite} representation of ratio functions. And we make these results available in the ancillary files to this work's submission to the \texttt{arXiv}.

\subsection{Complete Prescriptivity at Two Loops}

Recently, a complete basis $\mathfrak{B}$ of maximal-weight, planar, two-loop master integrals for four dimensional theories was described \cite{Bourjaily:2026icc}. In terms of this basis, any amplitude in any four-dimensional theory may be represented as a linear combination of these integrals up to terms of strictly sub-maximal (`transcendental') `weight':\\[-10pt]
\vspace{4pt}\eq{\fwbox{0pt}{\hspace{-24pt}\int\!\!\dbar^4\hspace{-1pt}\ell_1\dbar^4\hspace{-1pt}\ell_2\,\,\,\mathcal{A}(\ell_1,\ell_2)\equivL\sum_{\mathcal{I}_i\in\mathfrak{B}}\hspace{-2pt}\mathfrak{a}^i\hspace{-4pt}\int\!\!\dbar^4\hspace{-1pt}\ell_1\dbar^4\hspace{-1pt}\ell_2\,\,\,\mathcal{I}_i(\ell_1,\ell_2){+}\left(\text{{\small terms of sub-leading `weight'}}\right)\!.}\label{unitarity}\vspace{-4pt}}
In this basis, each master integral is individually pure/canonical and supports (has unit period on) a \emph{specific} compact contour of integration not supported by any other master integral in the basis; that is, these masters, viewed as differential forms on loop momenta, are the cohomological duals of a particular choice of eight-dimensional, compact cycles of integration:
\eq{\int\limits_{\Omega_j}\!\!\dbar^4\hspace{-1pt}\ell_1\dbar^4\hspace{-1pt}\ell_2\,\,\,\mathcal{I}(\ell_1,\ell_2)=\delta\indices{j}{i}\,.\vspace{-4pt}\label{prescriptivity}}
As such, the coefficients of amplitudes with respect to this basis are simply given by the corresponding periods of amplitudes:
\vspace{-4pt}\eq{\mathfrak{a}^i=\int\limits_{\Omega_i}\!\!\dbar^4\hspace{-1pt}\ell_1\dbar^4\hspace{-1pt}\ell_2\,\,\,\mathcal{A}(\ell_1,\ell_2)\,,\vspace{-5pt}}
without any need for integrand reduction or integration by parts (`IBP') identities/projections. Because the cycles to which the basis is dual were chosen to each involve components which put at least three propagators on-shell, these periods of amplitudes are \emph{leading singularities}: computable as \emph{on-shell functions} constructed from trees and without any reference to gauge redundancy. 

This summary holds for amplitudes in \emph{any} quantum field theory given a choice of master integrals satisfying the \emph{prescriptivity} condition of (\ref{prescriptivity}). In the case of maximally supersymmetric ($\mathcal{N}{=}4$) Yang-Mills theory in the planar limit (`sYM'), it is known that no contributions of sub-leading weight arise in (\ref{unitarity}), which is to say that all amplitudes in this theory lie within the \emph{span} of these masters. 

\subsection{All Two-Loop Cuts Supported by Amplitudes in sYM}\label{two_loop_cuts_classified}

\begin{figure}[t]$$\begin{array}{@{}c@{}}
\fwbox{0pt}{\fwboxL{35pt}{\;\text{finite:\,}}\fwboxL{395pt}{\hspace{2pt}\left\{\rule{0pt}{30pt}\right.\hspace{1pt}\fwbox{75pt}{\hspace{-8pt}\fig{-2.25pt}{1}{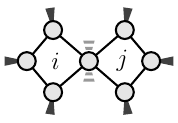}},\fwbox{70pt}{\hspace{-8pt}\fig{-2.25pt}{1}{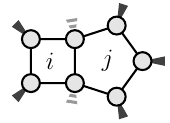}},\fwbox{65.5pt}{\fig{-2.25pt}{1}{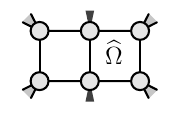}},\fwbox{65.5pt}{\fig{-2.25pt}{1}{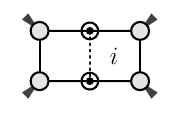}},\fwbox{67.5pt}{\hspace{-3pt}\fig{-2.25pt}{1}{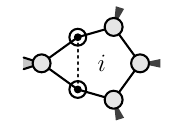}}\hspace{-5.5pt}\left.\rule{0pt}{30pt}\right\}
}}\\[-9pt]
\fwbox{0pt}{\fwboxL{75pt}{\;\text{$\log^2$-divergent:\,}}\fwboxL{355pt}{\hspace{6pt}\left\{\rule{0pt}{30pt}\right.\hspace{1pt}\fwbox{70.5pt}{\hspace{-7.5pt}\fig{-2.25pt}{1}{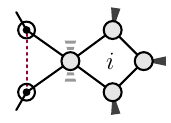}}\hspace{2pt},\fwbox{65pt}{\fig{-2.25pt}{1}{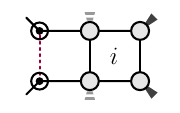}},\fwbox{67.5pt}{\hspace{-3pt}\fig{-2.25pt}{1}{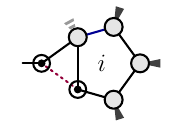}},\fwbox{67.5pt}{\hspace{-3pt}\fig{-2.25pt}{1}{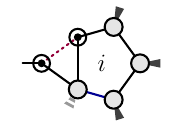}}\hspace{-6.5pt}\left.\rule{0pt}{30pt}\right\}
}}\\[-9pt]
\fwbox{0pt}{\fwboxL{75pt}{\;\text{$\log^4$-divergent:\,}}\fwboxL{355pt}{\hspace{6pt}\left\{\rule{0pt}{30pt}\right.\hspace{1pt}\fwbox{50.5pt}{\hspace{-1pt}\fig{-2.25pt}{1}{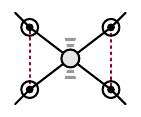}},\fwbox{55pt}{\hspace{-5pt}\fig{-2.25pt}{1}{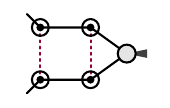}},\fwbox{55pt}{\hspace{-5pt}\fig{-2.25pt}{1}{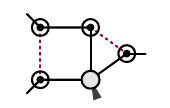}},\fwbox{67.5pt}{\hspace{-3pt}\fig{-2.25pt}{1}{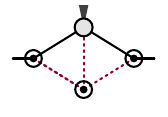}},\fwbox{67.5pt}{\hspace{-3pt}\fig{-2.25pt}{1}{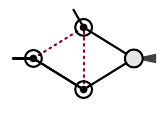}}\hspace{-5.5pt}\left.\rule{0pt}{30pt}\right\}
}}\end{array}$$\vspace{-30pt}\caption{The complete list of two-loop leading singularities supported by amplitudes in planar sYM. {Note}: these figures represent dihedral equivalence classes; also the label `$\log^4$-divergent' should be understood as \emph{at most} $\log^4$-divergent---as the second and last of these cuts are in fact associated with mere $\log^2$-divergences.}\label{all_two_loop_leading_singularities_of_sym}\vspace{-10pt}\end{figure}

At two loops, amplitudes in sYM theory support periods on contours associated with the following on-shell diagrams.\footnote{It may surprise the reader to learn that until undertaking the present work, the author was not even aware that some of these contours had non-vanishing support in sYM.} These are naturally organized according to whether or not the cut loop momenta are in a region associated with a source of infrared divergence (soft or soft-collinear regions necessarily living along the Feynman contour of loop integration), and to what degree. These are illustrated in \mbox{figure~\ref{all_two_loop_leading_singularities_of_sym}}. 

In terms of colour-stripped partial (ordered) amplitudes, the only leading singularities not encountered at lower loops are the first three `finite' ones drawn in \mbox{figure~\ref{all_two_loop_leading_singularities_of_sym}}:
\vspace{-4pt}\eq{\fwbox{0pt}{\fwbox{395pt}{\hspace{2pt}\left\{\rule{0pt}{30pt}\right.\hspace{5pt}\fwbox{75pt}{\hspace{-8pt}\fig{-2.25pt}{1}{404_ls}},\fwbox{70pt}{\hspace{-8pt}\fig{-2.25pt}{1}{413_ls}},\fwbox{65.5pt}{\fig{-2.25pt}{1}{313_ls_e}}\hspace{-1.5pt}\left.\rule{0pt}{30pt}\right\}}}\vspace{-4pt}}
while the rest are either associated directly with a (functionally) `trivial' \emph{decoration} on a one-loop leading singularity
\vspace{-12pt}\eq{\left\{\rule{0pt}{26pt}\right.\fwbox{37.5pt}{\fig{-2.25pt}{1}{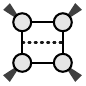}},\fwbox{37.5pt}{\fig{-2.25pt}{1}{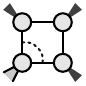}},\fwbox{62.5pt}{\hspace{-2pt}\fig{-2.25pt}{1}{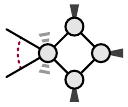}},\fwbox{45pt}{\hspace{-1pt}\fig{-2.25pt}{1}{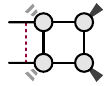}},\fwbox{40pt}{\fig{-2.25pt}{1}{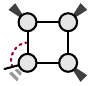}},\fwbox{40pt}{\fig{-2.25pt}{1}{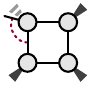}}\hspace{-3pt}\left.\rule{0pt}{26pt}\right\}\vspace{-14pt}\label{one_loop_decorations}}
or---in the case of all those associated with a $\log^4$-divergence---a decoration of the tree. That is, as periods of loop amplitudes, these decorated leading singularities are functionally identical to lower-loop leading singularities. Of course, their dressing by colour factors would obviously be distinct; examples of this structure were encountered in, for example, the works of \cite{Bourjaily:2018omh,Bourjaily:2019iqr,Bourjaily:2019gqu,Bourjaily:2021iyq,Bourjaily:2021hcp} (see also~\cite{Bourjaily:2025hvq,Bourjaily:2026whq}). 

In the basis of master integrals described in \cite{Bourjaily:2026icc}, each of contours in \mbox{figure~\ref{all_two_loop_leading_singularities_of_sym}} is paired with a unique master integral, whose period on this contour has been normalized to be 1. Thus, to represent amplitudes of sYM in this basis, we need only multiply each coefficient by its corresponding master integral and sum over all (inequivalent) contributions. Technically speaking, the primary challenge in implementing this representation is one of book-keeping: ensuring that appropriate indices and signs have been matched between integrals and coefficients across terms. We have implemented this in the ancillary files for this work, which we describe in \mbox{section~\ref{sec_ancillary_files}}.\\[-6pt]

It is worth commenting on the fact that the (vast) majority of master integrals described in \cite{Bourjaily:2026icc} were defined for contours on which all amplitudes \emph{of sYM} vanish. This is obviously the case for those defined to match contours involving one or both loop momenta sent to infinity such as
\vspace{-8pt}\eq{\fig{-2pt}{1}{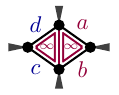}\vspace{-8pt}\label{eg_cut_at_infinity}}
---as amplitudes in sYM are dual-conformally invariant. But it is much less obvious that amplitudes vanish on a contour such as,
\vspace{-5pt}\eq{\fig{-2.25pt}{1}{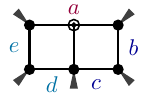}\vspace{-5pt}\label{eg_spurious_cut}}
for example. Amplitudes in pure YM \emph{do} have support on cuts such as (\ref{eg_cut_at_infinity}); but we (strongly) suspect that in no quantum field theory do amplitudes have support on contours such as (\ref{eg_spurious_cut}). As such, many of the master integrals in \cite{Bourjaily:2026icc} not relevant to sYM will still be required for pure YM, or other four-dimensional theories.\\[-6pt]  

Even though relatively few integrals in the basis of \cite{Bourjaily:2026icc} are needed to represent amplitudes in sYM, there is considerable weight lifted by the entire basis: every integral in the basis supporting a contour \emph{relevant} to amplitudes also supports a large number of irrelevant other periods of integration; it is only their \emph{completeness} and \emph{orthogonality} (`prescriptivity') that \emph{ensures} that once we have matched some loop integrand on each of the contours used to define the masters, it must agree everywhere (up to terms of lower transcendental weight).

For example, while none of the masters are dual-conformally invariant, the dual-conformal invariance of the amplitude is ensured by the completeness of the basis. The vanishing of the amplitude on any cut which sends one or both loop momentum to infinity is guaranteed to follow via completeness of the basis; it will appear as an (often highly non-trivial) instance of the `global residue theorem'.\\[-6pt]

\subsection{Organization and Outline}\label{sec_organization}

This work is organized as follows. In \mbox{section~\ref{sec_review_of_local_integral_representations}} we review various existing representations of amplitudes in sYM at the integral and integrand-levels. In particular, we discuss the scalar box expansion and the no-triangle hypothesis in \mbox{section~\ref{sec_review_conformality_at_one_loop}}, and describe how this may be upgraded to a manifestly (dual-)conformal representation of amplitudes at one loop in \mbox{section~\ref{sec_manifesting_conformality_one_loop}}. We then illustrate the incompleteness of such representations by reviewing the chiral box expansion of \cite{Bourjaily:2013mma} in \mbox{section~\ref{sec_complete_prescriptivity_one_loop}}. 

In \mbox{section~\ref{review_of_conformal_a2}} we revisit the construction of two loop amplitudes in planar sYM at the integrand-level as described in \cite{Bourjaily:2015jna}, both for the sake of review and to sharpen the role of the completely prescriptive, novel representation described here. 

Our main result is outlined in \mbox{section~\ref{sec_complete_prescriptive_a2}}, and in \mbox{section~\ref{sec_organization_of_terms_for_new_a2}} we present a complete enumeration of the terms that contribute to two-loop amplitudes when represented in terms of the basis described in \cite{Bourjaily:2026icc}. 

In \mbox{section~\ref{sec_ratio_functions}} we review the general structure of the perturbative ratio functions constructed from perturbative amplitudes involving massless particles, and in \mbox{section~\ref{sec_local_finiteness_of_r2}} we show that complete prescriptivity at two loop ensures that all ratio functions are \emph{manifestly} infrared-finite in this representation---that is, all ratio functions may be computed without reference to regularization or schemes in four dimensions. 

As is often the case in amplitudes research, the many subtleties of formulae (including whatever peculiar conventions must be chosen) are best clarified through explicit, human-usable, machine code for expressions. These are provided as ancillary files to this work's submission files to the \texttt{arXiv}, and consist of a number of computational tools and reference formulae encoded in a \textsc{Mathematica} package \texttt{prescriptive\uscore amps.m}, together with an illustration \texttt{Notebook} which walks the user through much of the primary functionality provided. The main features of this ancillary code is described in \mbox{section~\ref{sec_ancillary_files}}. 

Finally, we conclude our work with a number of applications and future investigations in \mbox{section~\ref{sec_conclusions}}.

\newpage

\section[{Review: Prescriptive Representations of Amplitudes in sYM}]{Review: Representations of Amplitudes in sYM}\label{sec_review_of_local_integral_representations}\vspace{-4pt}

\subsection{Conformal vs.\ Non-Conformal Representations at One Loop}\label{sec_review_conformality_at_one_loop}

In any $d$-dimensional quantum field theory at one loop, master integrals may be chosen to consist of  \emph{scalar} Feynman integrals with at most $d$ propagators and (a large number of) masters that vanish upon loop integration. In (unregulated) four-dimensional theories, this means that (the integral over loop momentum of) any scattering amplitude may be represented in terms of boxes, triangles, and bubbles,
\eq{\int\limits_{\r{\ell}\in\mathbb{R}^{4}}\!\!\!\!\dbar^{\hspace{-1pt}4}\hspace{-1pt}\r{\ell}\,\,\mathcal{A}^{(1)}(\r{\ell})=\sum_{a,b,c,d}\mathfrak{a}_{a,b,c,d}\,I_{a,b,c,d}{+}\sum_{a,b,c}\mathfrak{a}_{a,b,c}\,I_{a,b,c}{+}\sum_{a,b}\mathfrak{a}_{a,b}\,I_{a,b},\vspace{-5pt}\label{one_loop_scalar_expansion}}
with loop-independent coefficients determined by integral reduction or (`generalized') unitarity \cite{Bern:1994zx,Bern:1994cg,Britto:2004nc} (see also \cite{Drummond:2008bq,Anastasiou:2006jv}). Here, 
\eq{I_{a,\ldots,d}\,\equivR\hspace{-4pt}\int\limits_{\r{\ell}\in\mathbb{R}^{4}}\!\!\!\!\dbar^{\hspace{-1pt}4}\hspace{-1pt}\r{\ell}\,\,\mathcal{I}_{a,\ldots,d}(\r{\ell}),}
with $\dbar\equivR d/(2\pi)$ (by analogy with $\hbar$) so that `residues' are in fact directly period integrals of the measure. 

For the sake of concreteness and convenience, we choose the loop-independent normalizations of these integrals in terms of the invariants $s_{a\,b}\equivR(p_a{-}p_b)^2$ as follows:\\[-10pt]
\eq{\begin{split}&\mathcal{I}_{a,b,c,d}\equivR\frac{1}{2}\frac{\sqrt{(s_{a\,c}\,s_{b\,d}{-}s_{a\,b}\,s_{c\,d}{-}s_{b\,c}\,s_{b\,d})^2{-}4\,s_{a\,b}\,s_{b\,c}\,s_{c\,d}\,s_{d\,a}}}{(\r{\ell}{-}p_a)^2(\r{\ell}{-}p_b)^2(\r{\ell}{-}p_c)^2(\r{\ell}{-}p_d)^2},\\
&\mathcal{I}_{a,b,c\phantom{,d}}\equivR\frac{1}{2}\frac{\sqrt{(s_{a\,b}{+}s_{b\,c}{-}s_{c\,a})^2{-}4\,s_{a\,b}\,s_{b\,c}}}{(\r{\ell}{-}p_a)^2(\r{\ell}{-}p_b)^2(\r{\ell}{-}p_c)^2},\hspace{26pt}\text{and}\hspace{26pt}\mathcal{I}_{a,b}\equivR\frac{1}{2}\frac{1}{(\r{\ell}{-}p_a)^2(\r{\ell}{-}p_b)^2}.
\end{split}\label{unit_scalar_ints_one_loop}}
When so normalized, the box integrals are (dual-)\emph{conformally invariant} with respect to the (`dual') momenta $p_a$ appearing in the loop-dependent propagators. It will be useful later to also express these integrands in embedding/momentum-twistor space, in which they would be written (see \emph{e.g.}~\cite{Hodges:2009hk,Bourjaily:2019exo,Bourjaily:2020qca} for details of this formalism)
\eq{\fwbox{0pt}{\hspace{-25pt}\mathcal{I}_{a,b,c,d}\equivR\frac{\frac{1}{2}\Delta[a,\!b,\!c,\!d]\phantom{\frac{1}{2}}}{\x{\r{\ell}}{a}\x{\r{\ell}}{b}\x{\r{\ell}}{c}\x{\r{\ell}}{d}},\hspace{2pt}\mathcal{I}_{a,b,c,\inf}\equivR\frac{\frac{1}{2}\Delta[a,\!b,\!c,\!\inf]\phantom{\frac{1}{2}}}{\x{\r{\ell}}{a}\x{\r{\ell}}{b}\x{\r{\ell}}{c}\x{\r{\ell}}{\inf}},\hspace{2pt}\mathcal{I}_{a,b,\inf}\equivR\frac{\frac{1}{2}\x{a}{\inf}\x{b}{\inf}\phantom{\frac{1}{2}}}{\x{\r{\ell}}{a}\x{\r{\ell}}{b}\x{\r{\ell}}{\inf}^2}}\label{unit_scalar_ints_one_loop_in_mt}}
where
\eq{\Delta[a,\!b,\!c,\!d]\equivR\sqrt{(\x{a}{c}\x{b}{d}{-}\x{a}{b}\x{c}{d}{-}\x{b}{c}\x{d}{a})^2{-}4\,\x{a}{b}\x{b}{c}\x{c}{d}\x{d}{a}}\,.}
(Notice that these integrals involve a \emph{conventionally-defined} choice of overall sign.)\\[-6pt]

%\\[-6pt]

Early computations of amplitudes in gauge theory led to the discovery that amplitudes in maximal supersymmetric Yang-Mills theory (and supergravity) satisfy the `no triangle hypothesis' \cite{Bern:1993qk,Bern:1998ug,Britto:2004nc,Cachazo:2008dx,Bern:2005bb,BjerrumBohr:2006yw}: that all triangle and bubble coefficients in (\ref{one_loop_scalar_expansion}) vanish, and (integrated) amplitudes may be represented entirely in terms of boxes alone. As we will see, this fact is closely tied to the \emph{dual} conformal invariance of the theory---and not merely its maximal `transcendentality'. 

The coefficients of boxes in (\ref{one_loop_scalar_expansion}) may be determined directly in terms of trees via generalized unitarity: consider a contour of integration which `cuts' (encircles---as a residue---the vanishing of) four propagators. There are generally two solutions to the cut equations $(\r{\ell}{-}p_a)^2{=}\cdots{=}(\r{\ell}{-}p_d)^2{=}0$; let's denote them $\Omega^i_{a,b,c,d}$.  Such a contour in field theory would select all Feynman graphs with a topology including these four propagators
\vspace{-10pt}\eq{\int\limits_{\Omega_{a\,b\,c\,d}^{i}}\!\!\!\!\dbar^{\hspace{-1pt}4}\hspace{-1pt}\r{\ell}\,\,\mathcal{A}^{(1)}(\r{\ell})\;\equivL\;\mathfrak{a}_{a,b,c,d}^{i}\bigger{\;\Leftrightarrow\;}\fig{-2.25pt}{1}{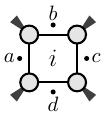}\,\vspace{-10pt}}
and as all possible tree-level graphs may appear at each corner, this coefficient must be computable in terms of \emph{tree}-amplitudes at the vertices evaluated with the on-shell, internal momentum specified by the cut (for each of its two solutions). These periods of amplitudes are called \emph{leading singularities} or `on-shell functions' in the parlance of \cite{ArkaniHamed:book} (see also \cite{Cachazo:2008vp,ArkaniHamed:2009dn,Ashok:2010ie,Kaplan:2009mh}); they may be determined without any reference to gauge redundancy or even virtuality.  

Consider integrating the RHS of (\ref{one_loop_scalar_expansion}) on such a contour. There will be only one term which is non-vanishing on such a contour: the scalar box $\mathcal{I}_{a,b,c,d}$. Our normalization in (\ref{unit_scalar_ints_one_loop}) was chosen so that
\vspace{-5pt}\eq{\int\limits_{\Omega_{a,b,c,d}^{\text{even}}}\!\!\!\!\dbar^{\hspace{-1pt}4}\hspace{-1pt}\r{\ell}\,\,\mathcal{I}_{a,b,c,d}(\r{\ell})\equivR\hspace{-5pt}\int\limits_{\Omega_{a,b,c,d}^{1}}\!\!\!\!\dbar^{\hspace{-1pt}4}\hspace{-1pt}\r{\ell}\,\,\mathcal{I}_{a,b,c,d}(\r{\ell})\;\;{-}\hspace{-5pt}\int\limits_{\Omega_{a,b,c,d}^{2}}\!\!\!\!\dbar^{\hspace{-1pt}4}\hspace{-1pt}\r{\ell}\,\,\mathcal{I}_{a,b,c,d}(\r{\ell})=1\,.\vspace{-5pt}\label{scalar_box_even_contour_defined}}
Notice that the `even' contour is defined as the \emph{difference} between the periods on the two cut solutions, because their \emph{sum} must vanish due to Cauchy's theorem. Matching the left and right hand sides of (\ref{one_loop_scalar_expansion}) (with an appropriate choice of signs for the on-shell functions involved), we arrive at:
\eq{\int\limits_{\r{\ell}\in\mathbb{R}^{4}}\!\!\!\!\dbar^{\hspace{-1pt}4}\hspace{-1pt}\r{\ell}\,\,\mathcal{A}^{(1)}(\r{\ell})=\sum_{a,b,c,d}\big(\mathfrak{a}^1_{a,b,c,d}{+}\mathfrak{a}^2_{a,b,c,d}\big)\,I_{a,b,c,d}\,.\label{scalar_box_expansion}}

\subsubsection{Manifestly (Dual) Conformally Invariant Amplitude Integrands}\label{sec_manifesting_conformality_one_loop}

Recall that the discovery of dual conformal invariance \cite{Drummond:2006rz} (and much else) of amplitudes in sYM followed from the observation that this was a symmetry of the \emph{unintegrated} \emph{integrands} appearing in (\ref{scalar_box_expansion}), and also a symmetry of their coefficients. This symmetry was also noted at strong coupling in  \cite{Alday:2007hr}, and led to the remarkable connection between null-polygonal Wilson loops and scattering amplitudes---see \emph{e.g.}~\cite{Alday:2007hr,Alday:2007he,Brandhuber:2007yx,Drummond:2007aua,Drummond:2007cf,Drummond:2007bm,Drummond:2008aq,Bern:2008ap,Alday:2009yn}---and the emergence of the Yangian \cite{Drummond:2009fd,Drummond:2010uq,Drummond:2010qh,Beisert:2010gn}.

But is dual-conformal invariance a true \emph{symmetry} of amplitudes? Surely not: amplitudes involving massless particles are \emph{not even observables}, being mired as they are by real (and important) infrared divergences. The infrared divergences of amplitudes obfuscate any the meaning of \emph{their} conformal invariance, as regularization of these divergences typically (if not necessarily---see \emph{e.g.}~\cite{Bourjaily:2013mma}) breaks conformal invariance. 

So what is to be understood from the conformal invariance of (\ref{scalar_box_expansion}) when viewed at the \emph{integrand-level}? This is not immediately obvious, because (\ref{scalar_box_expansion}) is \emph{not} an equality at the integrand-level (beyond four particles):
\eq{\mathcal{A}^{(1)}(\r{\ell})\neq\sum_{a,b,c,d}\big(\mathfrak{a}^1_{a,b,c,d}{+}\mathfrak{a}^2_{a,b,c,d}\big)\,\mathcal{I}_{a,b,c,d}.\vspace{-10pt}}
The easiest way to see this inequality is by noting that the two cuts of field theory, $\mathfrak{a}^{1,2}$, are \emph{never} equal beyond four particles, while the scalar box's residues are always equal in magnitude (and opposite in sign). As such, we may only conclude that 
\eq{\hspace{-20pt}\mathcal{A}^{(1)}(\r{\ell})=\hspace{-4pt}\sum_{a,b,c,d}\hspace{-4pt}\big(\mathfrak{a}^1_{a,b,c,d}{+}\mathfrak{a}^2_{a,b,c,d}\big)\,\mathcal{I}_{a,b,c,d}{+}\big(\text{terms vanishing on Feynman contour}\big).\hspace{-16pt}\label{up_to_vanishing_one_loop}}
To be clear, these `vanishing terms' are \emph{not} cohomologically trivial: they do not vanish due to integration-by-parts identities. Rather, they involve independent elements of cohomology that just happen to integrate to zero over the Feynman contour of loop integration $\r{\ell}\!\in\!\mathbb{R}^4$. 

So what kinds of terms may appear in this `vanishing' part of the loop integrand? It turns out that the \emph{vast} majority of master Feynman integrals (may be chosen to) vanish upon loop integration; but this fact is surprisingly non-trivial as it is by no means obvious that arbitrary loop-dependence in the numerator of any Feynman graph may be reduced to \emph{just} \emph{scalars} and those tensor integrals conspiratorially defined to vanish upon loop integration. At any rate, it is not at all clear that the no-triangle hypothesis (let alone dual-conformality) will be a feature of amplitude integrands once the vanishing terms are included. But it is nevertheless true.\\[-6pt]

\addtocontents{toc}{\protect~\\[-34.5pt]}
\paragraph{Resolving a Tension between Conformality and Chirality}~\\[-16pt]

It's not hard to see that there are conformal integrals at one loop which vanish upon Feynman loop integration. A classic result of Passarino and Veltman \cite{Passarino:1978jh} (see also \cite{tHooft:1978xw,vanNeerven:1983vr}) that any 4-dimensional loop integral with at 6 or more propagators is reducible (as a rational function---not using integration by parts identities) into those with 5 or fewer; as the scalar boxes represent all (potentially) conformal integrals with 4 propagators, we are led to consider also `pentagon' integrals involving 5 propagators. 

Conformality at one loop requires that any integral with $(4{+}p)$ propagators must involve at least $p$ powers of inverse propagators in the numerator. The span of inverse propagators is $(d{+}2)$-dimensional, and so the loop dependence of any pentagon integral can be expressed as a linear combination of the five boxes that comprise its `contact terms' together with one (unique up to its loop-independent scaling) element in the complement of these five terms. (See \emph{e.g.}~\cite{Bourjaily:2020qca} for more thorough discussions of this accounting.) Up to a choice of overall scaling then, these missing masters are essentially unique. Once properly normalized (so-as to be pure (and hence also conformal)), they may be expressed in terms of four-momenta as
\eq{\begin{split}~\\[-48pt]\mathcal{I}^{\text{odd}}_{a,b,c,d,e}(\r{\ell})\equivR\frac{(\r{\ell}{-}p_a)^\mu(p_a{-}p_b)^{\nu}(p_b{-}p_c)^\rho(p_c{-}p_d)^\sigma(p_d{-}p_e)^{\tau}(\r{\ell}{-}p_e)^{\upsilon}\mathrm{tr}\big(\gamma^5\gamma_{\mu}\gamma_{\nu}\gamma_{\rho}\gamma_{\sigma}\gamma_{\tau}\gamma_{\upsilon}\big)}{2(\r{\ell}{-}p_a)^2(\r{\ell}{-}p_b)^2(\r{\ell}{-}p_c)^2(\r{\ell}{-}p_d)^2(\r{\ell}{-}p_e)^2}\\[-30pt]\end{split}\vspace{-20pt}\label{explicit_odd_pentagon}}\vspace{-5pt}
where
\eq{\begin{split}\mathrm{tr}\big(\gamma^5\gamma_{\mu}\gamma_{\nu}\gamma_{\rho}\gamma_{\sigma}\gamma_{\tau}\gamma_{\upsilon}\big)=&\,i(\eta_{\rho\mu}\,\epsilon_{\upsilon\tau\sigma\nu}{-}\eta_{\mu\nu}\,\epsilon_{\rho\sigma\tau\upsilon}{+}\eta_{\nu\rho}\,\epsilon_{\sigma\tau\upsilon\mu}\\&{-}\eta_{\sigma\upsilon}\,\epsilon_{\tau\mu\nu\rho}{+}\eta_{\sigma\tau}\,\epsilon_{\upsilon\mu\nu\rho}{-}\eta_{\tau\upsilon}\,\epsilon_{\mu\nu\rho\sigma}\big);\end{split}}
alternatively, this integrand may be written in embedding space \cite{Bourjaily:2019exo} or momentum-twistor \cite{Hodges:2009hk} variables (more simply) as
\eq{\mathcal{I}^{\text{odd}}_{a,b,c,d,e}(\r{\ell})\equivR\frac{\epsilon(\r{\ell},a,b,c,d,e)}{\x{\r{\ell}}{a}\x{\r{\ell}}{b}\x{\r{\ell}}{c}\x{\r{\ell}}{d}\x{\r{\ell}}{e}}\,.\label{odd_pentagon}}

These parity-odd, vanishing pentagon integrals are individually conformally invariant and obviously cohomologically distinct from the scalar boxes; and so they are natural candidates to span the `vanishing terms' appearing in (\ref{up_to_vanishing_one_loop}). Thus, the question of the conformality of amplitudes at the integrand-level amounts to the question of whether or not amplitudes' integrands live within the space of integrands \emph{spanned} by scalar boxes and these parity-odd pentagons. The answer turns out to be yes, but there is an interesting reason why amplitude integrands have not been explicitly represented in this way in the literature---even in the case of planar sYM, for which many other integrand-level expansions are well known and widely available (see \emph{e.g.}~\cite{ArkaniHamed:2010gh,Bourjaily:2013mma}). 

The obstruction to any discussion of \emph{the} coefficient of any \emph{particular} parity-odd pentagon integral is the simple fact that they are not linearly (nor cohomologically) independent of one another. While there are $\binom{n}{5}$ pentagons to discuss, they satisfy $\binom{n{-}1}{5}$ linear relations. As such, any choice of independent pentagons would \emph{seem} to be arbitrary, and its coefficient obviously basis-dependent. Moreover, even if such a basis were chosen, each pentagon integral would have support on a variety of box-cuts, rendering the determination of their coefficients to appear to be an unfortunate exercise in linear algebra.\\[-6pt]

It turns out that we can nicely resolve both concerns by first allowing ourselves to forego conformal invariance (if only temporarily). Rather than conformal, parity-odd pentagon integrals, consider instead the non-conformal, parity-odd boxes: 
\eq{\mathcal{I}^{\text{odd}}_{a,b,c,d}(\r{\ell})\equivR\frac{2i\,(p_a{-}p_b)^{\mu}(\r{\ell}{-}p_b)^\nu(\r{\ell}{-}p_c)^\rho(p_c{-}p_d)^{\sigma}\epsilon_{\mu\nu\rho\sigma}}{(\r{\ell}{-}p_a)^2(\r{\ell}{-}p_b)^2(\r{\ell}{-}p_c)^2(\r{\ell}{-}p_d)^2},}
which, in embedding space or momentum twistor variables becomes
\eq{\mathcal{I}^{\text{odd}}_{a,b,c,d,\inf}(\r{\ell})\equivR\frac{\epsilon(\r{\ell},a,b,c,d,\inf)}{\x{\r{\ell}}{a}\x{\r{\ell}}{b}\x{\r{\ell}}{c}\x{\r{\ell}}{d}\x{\r{\ell}}{\inf}}\,.\label{odd_box}}
These integrals have been normalized so that 
\vspace{-5pt}\eq{\int\limits_{\Omega_{a,b,c,d}^{\text{odd}}}\!\!\!\!\dbar^{\hspace{-1pt}4}\hspace{-1pt}\r{\ell}\,\,\mathcal{I}^{\text{odd}}_{a,b,c,d,\inf}(\r{\ell})\equivR\hspace{-5pt}\int\limits_{\Omega_{a,b,c,d}^{1}}\!\!\!\!\dbar^{\hspace{-1pt}4}\hspace{-1pt}\r{\ell}\,\,\mathcal{I}^{\text{odd}}_{a,b,c,d,\inf}(\r{\ell})\;\;{+}\hspace{-5pt}\int\limits_{\Omega_{a,b,c,d}^{2}}\!\!\!\!\dbar^{\hspace{-1pt}4}\hspace{-1pt}\r{\ell}\,\,\mathcal{I}^{\text{odd}}_{a,b,c,d,\inf}(\r{\ell})=1\,.\vspace{-5pt}\label{odd_box_contour_defined}}
(Contrast this with (\ref{scalar_box_even_contour_defined}) for the scalar boxes above.)

Although the parity-odd boxes are not conformal, they are linearly and cohomologically independent (as evidenced by the fact that they are supported on distinct contours of integration). They outnumber the (independent) pentagons by $\binom{n{-}1}{3}$, but it is not hard to see that the parity-odd boxes \emph{span} the conformal pentagons:\\[-12pt]
\eq{\mathcal{I}^{\text{odd}}_{a,b,c,d,e}{=}\,\mathcal{I}^{\text{odd}}_{a,b,c,d,\inf}{+}\mathcal{I}^{\text{odd}}_{b,c,d,e,\inf}{+}\mathcal{I}^{\text{odd}}_{c,d,e,a,\inf}{+}\mathcal{I}^{\text{odd}}_{d,e,a,b,\inf}{+}\mathcal{I}^{\text{odd}}_{e,a,b,c,\inf}}
for \emph{any} choice of $\inf$.

Considering the periods defining the (parity-even) scalar and parity-odd boxes, (\ref{scalar_box_even_contour_defined}) and (\ref{odd_box_contour_defined}), it is not hard to see that the integral-level statement of the box expansion (\ref{scalar_box_expansion}) may be upgraded naturally to the \emph{integrand-level} statement that:
\vspace{-5pt}\eq{\mathcal{A}^{(1)}=\sum_{a,b,c,d}\Big[\mathfrak{a}^1_{a,b,c,d}\big(\mathcal{I}_{a,b,c,d}{+}\mathcal{I}^{\text{odd}}_{a,b,c,d,\inf}\big){+}\mathfrak{a}^2_{a,b,c,d}\big(\mathcal{I}_{a,b,c,d}{-}\mathcal{I}^{\text{odd}}_{a,b,c,d,\inf}\big)\Big].\label{non_conformal_chiral_box_expansion}\vspace{-5pt}}

In terms of this expansion, conformality of the integrand would be equivalent to the criterion that (\ref{non_conformal_chiral_box_expansion}) be independent of $\inf$. Whether in embedding space or in momentum-twistors, it is clear that $\inf$ plays a role equivalent to any propagator; and so, knowing that amplitude integrands in sYM are in fact conformal, we may restore \emph{manifest} conformality in (\ref{non_conformal_chiral_box_expansion}) simply by sending $\inf\!\mapsto\!\b{p_n}$---where $1/(\ell{-}\b{p_n})^2$ represents any one of the $n$ propagators, arriving at
\eq{\mathcal{A}^{(1)}=\sum_{a,b,c,d}\Big[\mathfrak{a}^1_{a,b,c,d}\big(\mathcal{I}_{a,b,c,d}{+}\mathcal{I}^{\text{odd}}_{a,b,c,d,\b{n}}\big){+}\mathfrak{a}^2_{a,b,c,d}\big(\mathcal{I}_{a,b,c,d}{-}\mathcal{I}^{\text{odd}}_{a,b,c,d,\b{n}}\big)\Big].\label{conformal_box_expansion}}

It is natural to be concerned that for boxes already involving $\b{n}$---that is, when $\b{n}\!\in\!\{a,b,c,d\}$---these integrands may involve a doubled propagator.  However, it is easy to see that the numerator in (\ref{odd_box}) vanishes trivially in this case. Moreover, when $\b{n}\!\notin\!\{a,b,c,d\}$, the integral becomes identical to that of an ordinary, parity-odd (conformal) pentagon (\ref{odd_pentagon}). Thus, sending $\inf\!\to\!\b{n}$ in fact picks out a preferred choice for $\binom{n{-}1}{4}$ parity-odd pentagons which form a linearly-independent basis.\\[-6pt]

As far as we know, the form of amplitudes at one loop in sYM represented at the integrand-level by (\ref{conformal_box_expansion}) has not appeared in the literature. It manifestly dual-conformally invariant term-by-term and entirely prescriptive. It arguably deserves to be called the `chiral box expansion'---although the authors of \cite{Bourjaily:2013mma} had their reasons to use this name for a rather different representation, which we should now describe.\\[-6pt]

\subsubsection{Complete Prescriptivity: Matching Every Cut at One Loop}\label{sec_complete_prescriptivity_one_loop}

Because amplitudes in sYM are conformal, only cuts involving physical propagators have support: any period defined by sending $\r{\ell}\!\to\!\infty$---or equivalently, cutting $`\x{\r{\ell}}{\inf}$'---must vanish. Are there any other contours on which amplitudes have support? 

Contours not involving $\inf$ may defined for any subset of three physical propagators for which at least two differ by a massless momentum. That is, for any triangle involving a massless leg. In such a case, cutting the two propagators differing by a massless momentum has two branches which differ by parity; the intersection of these two conditions, then defines a three-dimensional compact contour of integration which sends $\r{\ell}$ to be collinear with the inflowing momentum which separates them. Cutting any third propagator would then result in a compact, four-dimensional cycle of the form either 
\eq{\fig{-2.25pt}{1}{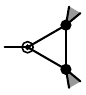}\quad\text{or}\quad\fig{-2.25pt}{1}{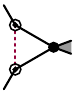}\label{composite_cuts_at_one_loop}}
depending on whether the third propagator differs from the first pair by a massless corner or not. In the first case of (\ref{composite_cuts_at_one_loop}), $\r{\ell}$ is merely \emph{collinear} with the massless corner and cuts the opposing propagator; in the second case, the loop momentum must be simultaneously collinear to both massless momenta---which in fact requires that the loop momentum between these legs must vanish. 

Both of these regions of loop momenta are \emph{real} and live along the physical Feynman contour of loop integration. As a result, these regions are responsible for \emph{infrared divergences} of loop amplitudes (or integrands) involving massless external particles. 

Notice that scalar box integrals involving at least one massless leg will always have support on such regions of internal loop momenta and therefore are divergent in the infrared. (For amplitudes involving fewer than 8 massless particles in (s)YM, these include \emph{every} scalar box integral.) Thus, the box expansion, whether viewed at the integrated-level (\ref{scalar_box_expansion}) or the integrand-level (\ref{conformal_box_expansion}) is littered by terms with infrared divergences. This is not obviously a bad thing: \emph{all} {amplitudes} involving massless in four dimensions are (\emph{importantly}) infrared-divergent.

While many terms in (\ref{scalar_box_expansion}) contribute to contours of both types in (\ref{composite_cuts_at_one_loop}), it turns out that no amplitudes \emph{ever}(?) have support on the former, while \emph{all} amplitudes have support on the latter. Moreover, as all infrared (and ultraviolet) divergences at one-loop must be proportional to the tree, the corresponding period of the amplitude on the soft-collinear contour will be given by the tree-amplitude:
\eq{\fig{-2.25pt}{1}{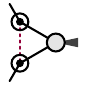}=\mathcal{A}_{\text{tree}}\equivL\mathcal{A}^{(0)}\,.\label{one_loop_soft_collinear_res}}
(The equality above refers only to the kinematic-dependent, `colour-stripped' part of the amplitude; the colour-\emph{dressed} leading singularity above involves distinct colour tensors.)

Viewed from the box expansion, the fact that summing over all box-coefficients which share three `adjacent' propagators (on either cut) gives the tree was the historic insight that gave rise to on-shell recursion \cite{BCF,BCFW} (see also \cite{Roiban:2004ix}) at tree-level, and ultimately beyond \cite{ArkaniHamed:2010kv,Bourjaily:2023apy}.\\[-6pt]

The scalar box expansion---or, at least its chiral manifestations at the integrand-level---will clearly match every leading singularity of an amplitude involving four propagators. But it fails to \emph{manifest} the fact that amplitudes \emph{are} supported on (\ref{one_loop_soft_collinear_res}) and also fails to manifest the fact that all(?) amplitudes \emph{vanish} on all contours involving a soft but not collinear internal momentum (the first cut shown in (\ref{composite_cuts_at_one_loop})). Thus, the box expansion involves both \emph{unnecessary} cancellations between and {unnecessarily} \emph{indirect} support upon regions responsible for infrared divergence.

To make manifest \emph{every} cut gets matched at the integrand-level, we should:\\[-20pt]
\begin{enumerate}
\item redefine the scalar boxes so that they vanish on any contour of (\ref{one_loop_soft_collinear_res});
\item add to our masters the scalar triangles involving massless legs.
\end{enumerate}\vspace{-7pt}
The first of these criteria can be achieved simply by subtracting scalar triangles of the appropriate topologies:
\eq{\mathcal{I}^{\text{finite}}_{a,b,c,d}(\r{\ell})\equivR\mathcal{I}_{a,b,c,d}(\r{\ell}){-}\sum\big(\text{triangle contact-terms involving massless legs}\big).\label{finite_boxes_defined}}
For example, if the corner between propagators involving $(\r{\ell}{-}p_{\t{a}})$ and $(\r{\ell}{-}p_{\t{b}})$ is massless and the others are massive (a `three-mass' box), then
\eq{\fwbox{0pt}{\fwboxL{430pt}{\x{\t{a}}{\t{b}}{=}0}}\fwbox{0pt}{\mathcal{I}^{\text{finite}}_{\t{a},\t{b},c,d,\inf}(\r{\ell})\equivR\mathcal{I}_{\t{a},\t{b},c,d}(\r{\ell}){-}\mathcal{I}_{\t{a},\t{b},c,\inf}(\r{\ell}){-}\mathcal{I}_{d,\t{a},\t{b},\inf}(\r{\ell}).}}
Not only do these subtractions ensure that no `conspiracies' arise among the cuts supported by field theory---whether cancelation in the case of collinear-but-not-soft divergences or BCF `magic' in the case of soft-collinear regions---but they also remove any and all infrared divergences any scalar box. Of course, these finite boxes are no longer conformally-invariant. 

For the second criterion, we note that any triangle with a collinear but \emph{not soft-collinear} divergence should have vanishing coefficient. For the $n$ `one-mass' triangles which support soft-collinear divergences (on which amplitudes have support), let us define
\vspace{-5pt}\eq{\mathcal{I}_{\text{div}}^{\hspace{2pt}\r{a}}\bigger{\;\Leftrightarrow\;}\fig{-2.25pt}{1}{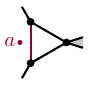}\qquad\text{and}\qquad\mathcal{I}^{(1)}_{\text{div}}\equivR\sum_{\r{a}\in[n]}\mathcal{I}^{\r{a}}_{\text{div}}\label{a1_divergences_defined}\vspace{-5pt}}

In terms of these, \emph{every} cut supported by field theory may be matched individually, and amplitude integrands may written
\vspace{-5pt}\eq{\begin{split}\mathcal{A}^{(1)}&=\sum_{a,b,c,d}\Big[\mathfrak{a}^1_{a,b,c,d}\big(\mathcal{I}^{\text{finite}}_{a,b,c,d}{+}\mathcal{I}^{\text{odd}}_{a,b,c,d,\inf}\big){+}\mathfrak{a}^2_{a,b,c,d}\big(\mathcal{I}^{\text{finite}}_{a,b,c,d}{-}\mathcal{I}^{\text{odd}}_{a,b,c,d,\inf}\big)\Big]{+}\mathcal{A}^{(0)}\,\mathcal{I}^{(1)}_{\text{div}}\\
&\equivL\Big[\sum_{\substack{a,b,c,d\\i}}\mathfrak{a}^{\hspace{0pt}i}_{a,b,c,d}\mathcal{I}^{\hspace{1pt}i}_{a,b,c,d}\Big]{+}\mathcal{A}^{(0)}\,\mathcal{I}^{(1)}_{\text{div}}\equivL\mathcal{A}^{(1)}_{\text{finite}}{+}\mathcal{A}^{(1)}_{\text{div}}\,.\\[-26pt]\vspace{-10pt}
\end{split}\label{a1_chiral_box_expansion}}
Notice that, following the conventions of \cite{Bourjaily:2013mma}, the \emph{chiral} box $\mathcal{I}^{i}_{a,b,c,d}$ is defined to be \emph{infrared finite} according to (\ref{finite_boxes_defined}) (as opposed to the integrals appearing in (\ref{non_conformal_chiral_box_expansion})---which, admittedly, are more evocative of the term `chiral box'.)  From our present perspective, we would call this \emph{completely} prescriptive representation, as every individual cut of field theory is matched by specific master integrals in the basis. 

We will have more to say about this representation in the context of finiteness in \mbox{section~\ref{sec_local_finiteness_of_r2}}.

\newpage
\subsection{Two-Loop Amplitudes Represented in a Basis of Conformal Masters}\label{review_of_conformal_a2}\vspace{2pt}

At two loops in planar sYM, local integral expressions were given in closed, manifestly dual-conformal form for all amplitudes at the integrand-level in \cite{Bourjaily:2015jna}. Such a representation should be viewed as the direct analogue of the form of amplitudes at one loop given by (\ref{conformal_box_expansion}). While this form captures \emph{some} of the infrared structure and matched \emph{some} of the individual cuts supported by field theory, a great deal of structure remained hidden, arising only indirectly. We do not wish to review that construction here, but briefly summarize its structure in order to better appreciate how it contrasts with the complete-prescriptivity of the form described in \mbox{section~\ref{sec_complete_prescriptive_a2}} below. 

Let us first discuss the space of master integrals at two loops which may be rendered dual-conformally invariant with appropriate normalization and loop dependence in their numerators. As described in \cite{Bourjaily:2020qca} (see also \cite{Bourjaily:2026icc}), this set of masters corresponds to those of the `box-power-counting' basis $\mathfrak{B}^4$ at two loops. These consist of masters of the form 
\vspace{-10pt}\eq{\left\{\rule{0pt}{28pt}\right.\hspace{-10pt}\fig{-2.25pt}{1}{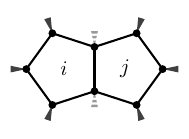},\hspace{-10pt}\fig{-2.25pt}{1}{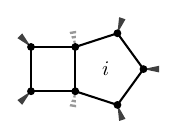}\hspace{-5pt},\hspace{-10pt}\fig{-2.25pt}{1}{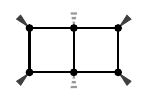}\hspace{-10pt}\left.\rule{0pt}{28pt}\right\}\vspace{-10pt}\label{conformal_two_loop_masters}}
where the indices $i,j\!\in\{1,2\}$ denote distinct choices of numerators. Notice that this basis does \emph{not} include the product of two scalar boxes---`kissing-box' integrals (which would also be conformal)---as these are in fact spanned already by the others:
\vspace{-10pt}\eq{\hspace{-5pt}\fig{-2.25pt}{1}{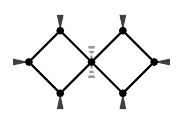}\hspace{-5pt}\subset\mathrm{span}\left\{\rule{0pt}{28pt}\right.\hspace{-10pt}\fig{-2.25pt}{1}{414_bare},\hspace{-10pt}\fig{-2.25pt}{1}{413_bare}\hspace{-5pt},\hspace{-10pt}\fig{-2.25pt}{1}{313_bare}\hspace{-10pt}\left.\rule{0pt}{28pt}\right\}\,.\vspace{-10pt}}
(It is perhaps surprising that the considerable annoyance of the \emph{over-completeness} of the parity-odd pentagons at one loop is entirely eliminated by the exclusion of the kissing boxes at two loops. (This niceness at two loops does not persist to higher loops, for which the over-completeness of box power-counting masters appears a difficult to resolve, general concern.))

Considering the enumeration of leading singularities \emph{with} support at two loops in \mbox{figure~\ref{all_two_loop_leading_singularities_of_sym}}, it is clear that many of these cannot be matched individually in any basis of conformal master integrals with the topologies of (\ref{conformal_two_loop_masters}). Moreover, it is easy to see that many of the leading singularities would in fact be supported by a considerable array of disparate masters.

To determine how amplitudes may be represented in terms of these masters for amplitudes, then, requires an active consideration of which subsets of cuts supported by field theory are \emph{particular} to a given master, and which may be matched indirectly through completeness of the basis. Let us briefly review the logic described in \cite{Bourjaily:2015jna} to derive a similar (if not identical) representation here.\\[-6pt]

One guiding principle is that the `finite cuts' should be matched by infrared finite master integrals where possible, $\log^2$-divergent cuts should be matched by $\log^2$-divergent integrals, and so-on. Of the masters given in (\ref{conformal_two_loop_masters}), all double-pentagons can be assigned numerators which render them infrared-finite, but any massless legs on the `box' part of the pentabox or in a corner of a double-box will render them infrared-divergent. The only finite integrals in the basis are the following:
\vspace{-10pt}\eq{\hspace{-2pt}\fig{-2.25pt}{1}{414_bare}\hspace{-2pt},\hspace{-10pt}\fig{-2.25pt}{1}{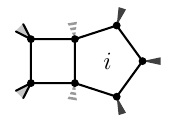}\hspace{-2pt},\hspace{-10pt}\fig{-2.25pt}{1}{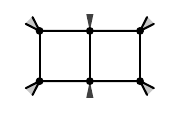}\hspace{-10pt},\hspace{-10pt}\fig{-2.25pt}{1}{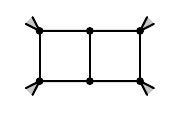}\hspace{-10pt},\hspace{-10pt}\fig{-2.25pt}{1}{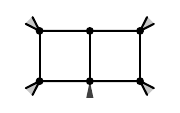}\vspace{-10pt}\label{finite_conformal_ints}}

To be clear, our conventions for labeling the momenta flowing into integrals and leading singularities are as follows:
\vspace{-5pt}\eq{\fig{-2pt}{1}{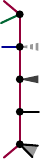}\begin{array}{@{}l@{$\;\;$}l@{}}
\hspace{2pt}&\text{\emph{\r{no}} momentum inflowing}\\[0pt]
\hspace{2pt}&\text{\emph{\r{any}} momentum, including \emph{none}}\\[0pt]
\hspace{2pt}&\text{\hspace{0pt}any \emph{\r{non-vanishing}} momentum}\\
\hspace{2pt}&\text{\hspace{0pt}any \emph{\r{massless}} momentum}\\[0pt]
\hspace{2pt}&\text{any \emph{\r{massive}} momentum}
\end{array}}

Which `finite' leading singularities should be assigned as coefficients of which of these integrals? Of those above, the only integrals supporting kissing-box leading singularities are the double pentagons, and so we must assign these as their coefficients. For the finite pentaboxes and double-boxes, there are fairly obvious choices of which subsets to choose; this leads to the \emph{finite-part} of the loop integrand being generated by the terms
\begin{align}\nonumber\\[-28pt]\mathcal{A}^{(2)}_{\substack{\text{conf.}\\\text{finite}}}=&\phantom{{+}}\Bigger{\sum}\hspace{-3pt}\fig{-2.25pt}{1}{404_ls}\!\times\!\hspace{-5pt}\fig{-2.25pt}{1}{414_bare}\hspace{-4pt}{+}\Bigger{\sum}\hspace{-5pt}\fig{-2.25pt}{1}{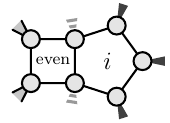}\!\times\!\hspace{-10pt}\fig{-2.25pt}{1}{413_c_int}\nonumber\\[-8pt]
&{+}\Bigger{\sum}\hspace{-10pt}\fig{-2.25pt}{1}{313_ls_e}\hspace{-14pt}\times\hspace{-14pt}\fig{-2.25pt}{1}{313_int_e}\hspace{-10pt}{+}\Bigger{\sum}\hspace{-10pt}\fig{-2.25pt}{1}{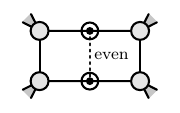}\hspace{-14pt}\times\hspace{-14pt}\fig{-2.25pt}{1}{313_c_int_c}\hspace{-10pt}\label{a2_finite_conformal}\\[-28pt]\nonumber\end{align}
Notice that this leaves \emph{implicit} a considerable fraction of the finite leading singularities, namely all those of 
\vspace{-6pt}\eq{\fwbox{85pt}{\fig{0pt}{1}{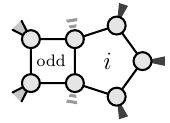}},\fwbox{75pt}{\hspace{-2pt}\fig{0pt}{1}{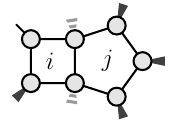}},\fwbox{65pt}{\fig{0pt}{1}{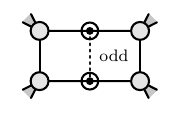}},
\fwbox{65pt}{\fig{0pt}{1}{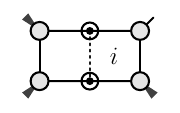}},\fwbox{65pt}{\fig{0pt}{1}{412_ls}}\label{implicit_finite_cuts}\vspace{-10pt}
}
It is a rather non-trivial exercise to show that once the subset of cuts in (\ref{a2_finite_conformal}) are matched correctly, all those of (\ref{implicit_finite_cuts}) are also matched correctly, although indirectly. (But this is, of course, \emph{guaranteed} by completeness in the basis combined with conformality of the amplitude integrand.)

Next, we'd like to match the $\log^2$-divergent parts of the amplitude, using masters that manifestly support a single soft-collinear divergence. These are easy to identify as integrals of one of two types:
\eq{\fig{-2.25pt}{1}{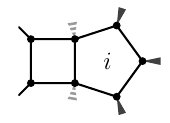}\quad\text{and}\quad\hspace{-10pt}\fig{-2.25pt}{1}{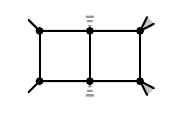}\hspace{-10pt}}
Considering the list of $\log^2$-divergent cuts supported in field theory, only the first of these can be used to match the kissing box-triangles, and only the second supports (half of) the double-box cuts. This pairing results in the representation of the $\log^2$-divergent part of the amplitude integrand as:
\vspace{-6pt}\eq{\mathcal{A}^{(2)}_{\substack{\text{conf.}\\\text{$\log^2\!\!\text{-div}$}}}\hspace{-3pt}=\Bigger{\sum}\hspace{-7pt}\fig{-2.25pt}{1}{403_d1_ls}\!\times\!\hspace{-10pt}\fig{-2.25pt}{1}{403_c_d1_int}\hspace{-4pt}{+}\Bigger{\sum}\hspace{-10pt}\fig{-2.25pt}{1}{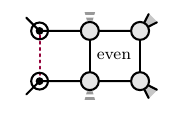}\hspace{-14pt}\!\times\!\hspace{-14pt}\fig{-2.25pt}{1}{313_c_d1_int}\label{a2_d1_conformal}\vspace{-10pt}}
As before, this representation leaves implicit a considerable number of cuts supported by field theory, namely all those of
\vspace{-12pt}\eq{\hspace{-10pt}\fig{-2.25pt}{1}{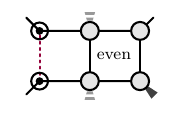}\hspace{-10pt},\hspace{-10pt}\fig{-2.25pt}{1}{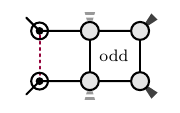}\hspace{-10pt},\hspace{-12pt}\fig{-2.25pt}{1}{412_d1_ls_1}\hspace{-8pt},\hspace{-12pt}\fig{-2.25pt}{1}{412_d1_ls_2}\hspace{-8pt},\hspace{-10pt}\fig{-2.25pt}{1}{312_d2_ls_1}\hspace{-10pt},\hspace{-7pt}\fig{-2.25pt}{1}{212_d2_ls_2}\hspace{-8pt}\vspace{-12pt}\nonumber}
And as before, it is rather non-trivial exercise to see that, given the cuts \emph{explicitly} matched by (\ref{a2_d1_conformal}), all those above will be matched correctly though indirectly. 

Finally, the only conformal integrals with a $\log^4$-infrared divergence are 
\vspace{-12pt}\eq{\fig{-2.25pt}{1}{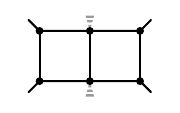}\vspace{-12pt}}
and these, of course, must of course be decorated by the tree---or, equivalently, any of the `$\log^4$-divergent' leading singularities in  \mbox{figure~\ref{all_two_loop_leading_singularities_of_sym}}. Thus, the final contribution to two-loop amplitude integrands is given by 
\vspace{-10pt}\eq{\mathcal{A}^{(2)}_{\substack{\text{conf.}\\\text{$\log^4\!\!\text{-div}$}}}\hspace{-3pt}=\Bigger{\sum}\hspace{-7pt}\fig{-2.25pt}{1}{303_d2_ls}\hspace{-8pt}\!\times\!\hspace{-8pt}\hspace{-10pt}\fig{-2.25pt}{1}{313_c_d2_int}\vspace{-10pt}}

Putting everything together, we find
\eq{\mathcal{A}^{(2)}_{\text{conf.}}=\mathcal{A}^{(2)}_{\substack{\text{conf.}\\\text{finite}}}+\mathcal{A}^{(2)}_{\substack{\text{conf.}\\\text{$\log^2\!\!\text{-div}$}}}+\mathcal{A}^{(2)}_{\substack{\text{conf.}\\\text{$\log^4\!\!\text{-div}$}}}\label{a2_conformal}}
where, importantly, the $\log^4$-divergent terms all directly involve the tree amplitude as their coefficients.\\[-6pt] 

\paragraph{Details on the Elliptic Cut Coefficients}\label{graph_on_elliptic_coefficients}~\\[-16pt]

Our representation here differs from that of \cite{Bourjaily:2015jna} mainly in the choice to include separately the elliptic and composite contributions in the finite part (\ref{a2_finite_conformal}); this has the advantage of making manifest the vanishing of coefficients for integrals of the last type in (\ref{finite_conformal_ints}), but at the cost of requiring coefficients that depend algebraically (as opposed to rationally) on the kinematics.\\[-6pt]

In light of the considerable interest in elliptic contributions to amplitudes at two and more loops (see \emph{e.g.}~\cite{Bourjaily:2017bsb,Bourjaily:2018ycu,Bourjaily:2018yfy,Bourjaily:2019hmc,Bourjaily:2022bwx}), it is worthwhile to sharpen the meaning of the coefficients assigned to the elliptic integrals above:
\vspace{-10pt}\eq{\fig{-2.25pt}{1}{313_ls_e}\vspace{-10pt}}
In the basis of masters described in \cite{Bourjaily:2026icc}, integrals were not diagonalized on elliptic contours for these topologies, but rather via \emph{evaluation} at specific points in loop-momentum space. This is analogous to the procedure used in \cite{Bourjaily:2015jna} to fix elliptic coefficients, and was discussed at greater length in \cite{Bourjaily:2021vyj}. In the language of \cite{Bourjaily:2020hjv}, if we define $\Omega_7$ to be a contour which puts the seven propagators of the double box on-shell (taking the difference of the two solutions to these conditions)
\vspace{-5pt}\eq{\begin{split}\mathfrak{d}(\r{\alpha})&\equivR\!\!\int\limits_{\Omega_7}\!\!\dbar^{\hspace{-1pt}4}\hspace{-1pt}\r{\ell_1}\dbar^{\hspace{-1pt}4}\hspace{-1pt}\r{\ell_2}\,\,\mathcal{A}^{(2)}(\r{\ell_1},\r{\ell_2})\\[-8pt]
&\equivL \frac{d\r{\alpha}}{y(\r{\alpha})}\mathfrak{db}_0{+}\big(\text{terms with simple poles from pentaboxes}\big).\\[-8pt]\end{split}\label{elliptic_seven_cut}}
where $y(\r{\alpha})^2$ is an irreducible quartic in $\r{\alpha}$; hence, this cut in field theory includes the geometry of an elliptic curve. In \cite{Bourjaily:2020hjv} it was shown that the period defined from (\ref{elliptic_seven_cut}) by integrating along any cycle of the elliptic curve would result in a Yangian-invariant on-shell function. If the master integrals of \cite{Bourjaily:2026icc} had been diagonalized against such periods, the coefficient of the double-box integrals would of course be this Yangian-invariant on-shell function, and, as argued in \cite{Bourjaily:2022tep}, any pentabox containing this elliptic sub-topology should be entirely polylogarithmic. 

However, such a prescription for the masters necessarily requires an incredible degree of algebraic (and transcendental) complexity into both the integrand basis and the space of coefficients. Because of this, it was decided that the basis of masters instead be chosen so that they vanish at some point $\r{\alpha^*}$ along this cut (\ref{elliptic_seven_cut}). This choice requires that the coefficient of double-box integrals be defined to be $\mathfrak{db}_0$---which, although not Yangian invariant, is much easier to compute and represent functionally.

As the reader can appreciate, while this representation of amplitudes manifests their conformal invariance, it is far from one which manifests each individual cut in field theory. Moreover, the fact that most of the infrared divergences are matched indirectly and implicitly across many separate contributions makes it much harder to manifest the infrared finiteness of observables such as the ratio function. 

\newpage
\section{\emph{Complete} Prescriptivity at Two Loops}\label{sec_complete_prescriptive_a2}

Like at one loop, the space of integrals with triangle power-counting is sufficient to individually match every cut supported by field theory,\mbox{figure~\ref{all_two_loop_leading_singularities_of_sym}}. Indeed, in the basis of masters described in \cite{Bourjaily:2026icc}, there is a unique integral normalized to have period 1 on each cut of field field theory; as such, representing amplitudes requires merely multiplying the right integrands by the right coefficients and summing.

\subsection{Organization of Contributions to Loop Amplitudes}\label{sec_organization_of_terms_for_new_a2}\vspace{4pt}

The contributions are naturally organized according to the strata in \mbox{figure~\ref{all_two_loop_leading_singularities_of_sym}}. In particular, \emph{all} of the finite leading-singularities may be individually matched by a \emph{manifestly finite} integral in the basis. Thus the finite part of two-loop amplitude integrands would be represented by
\begin{align}\nonumber\\[-28pt]\mathcal{A}^{(2)}_{\text{finite}}\equivR&\phantom{{+}}\Bigger{\sum}\hspace{-2pt}\fig{-2.25pt}{1}{404_ls}\hspace{-2pt}\!\times\!\hspace{-5pt}\fig{-2.25pt}{1}{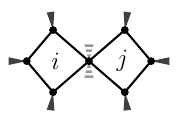}\hspace{-2pt}{+}\Bigger{\sum}\hspace{-5pt}\fig{-2.25pt}{1}{413_ls}\!\times\!\hspace{-10pt}\fig{-2.25pt}{1}{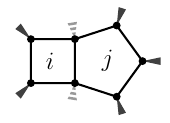}\nonumber\\[-16pt]
&{+}\Bigger{\sum}\hspace{-10pt}\fig{-2.25pt}{1}{313_ls_e}\hspace{-14pt}\times\hspace{-14pt}\fig{-2.25pt}{1}{313_int_e}\hspace{-10pt}{+}\Bigger{\sum}\hspace{-10pt}\fig{-2.25pt}{1}{313_ls_c}\hspace{-14pt}\times\hspace{-14pt}\fig{-2.25pt}{1}{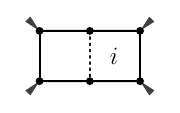}\nonumber\\[-16pt]
&{+}\Bigger{\sum}\hspace{-10pt}\fig{-2.25pt}{1}{412_ls}\hspace{-10pt}\times\hspace{-14pt}\fig{-2.25pt}{1}{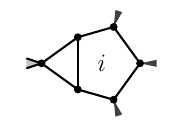}\label{a2_finite}\\[-28pt]\nonumber
\end{align}

Notice that, in contrast to the representation of (\ref{a2_conformal}), \emph{every} leading singularity is matched \emph{manifestly}, and each cut is the coefficient of a \emph{manifestly finite} (albeit non-conformal) master integral. For the elliptic coefficients, these are the same as for the conformal representation, and are determined as described in \ref{graph_on_elliptic_coefficients}.\\[-6pt]

Turning our attention to the $\log^2$-divergent cuts in \mbox{figure~\ref{all_two_loop_leading_singularities_of_sym}}, 
\vspace{-10pt}\eq{\left\{\rule{0pt}{30pt}\right.\fwbox{70.5pt}{\hspace{-7.5pt}\fig{-2.25pt}{1}{403_d1_ls}}\hspace{2pt},\fwbox{65pt}{\fig{-2.25pt}{1}{313_d1_ls}},\fwbox{67.5pt}{\hspace{-3pt}\fig{-2.25pt}{1}{412_d1_ls_1}},\fwbox{67.5pt}{\hspace{-3pt}\fig{-2.25pt}{1}{412_d1_ls_2}}\hspace{-6.5pt}\left.\rule{0pt}{30pt}\right\}.\label{a2_d1_redux}\vspace{-10pt}}
As before, for each leading singularity there is a unique master integrand of \cite{Bourjaily:2026icc} that supports this cut, and so the $\log^2$-contributions to amplitudes are as easy to enumerate as those of (\ref{a2_finite}) above. It is worthwhile to note that all these leading singularities are functionally identical to (after colour-stripping) a one-loop box coefficient, and to organize the terms in (\ref{a2_d1_redux}) first into the kissing topologies, 
\vspace{-6pt}\eq{\hspace{-140pt}\fwbox{65pt}{\fig{-2.25pt}{1}{4_ls_labelled}}\hspace{-10pt}\left\{\rule{0pt}{26pt}\right.\hspace{-5pt}\fwboxR{75pt}{\fig{-2.25pt}{1}{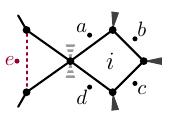}\hspace{-5pt}}\left.\rule{0pt}{26pt}\right\}\!\!\equivL\hspace{-5pt}\fwbox{65pt}{\fig{-2.25pt}{1}{4_ls_labelled}}\hspace{-5pt}\big(\mathcal{I}^{i}_{a,b,c,d}\!\times\!\mathcal{I}^{{\r{e}}}_{\text{div}}\big).\hspace{-100pt}%\mathcal{D}^{i,\hspace{4pt}\r{e}}_{\hspace{-2pt}{a},b,c,d}.\hspace{-100pt}
\vspace{-6pt}\label{two_loop_d_int_defined_1}}
the integrands of which are simply given by the products of the (finite) chiral boxes defined in (\ref{a1_chiral_box_expansion}) together with a soft-collinear divergent triangle integral (\ref{a1_divergences_defined}). 

For the remaining $\log^2$-divergent leading singularities, for reasons that will become clear when we consider ratio functions in the next section, it turns out to be worthwhile to organize these into groups which share overlapping regions of collinear divergence:
\vspace{-2pt}\eq{\hspace{-135pt}\fwbox{65pt}{\fig{-2.25pt}{1}{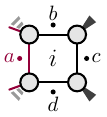}}\hspace{-10pt}\left\{\rule{0pt}{26pt}\right.\hspace{-5pt}\fwboxR{75pt}{\fig{-2.25pt}{1}{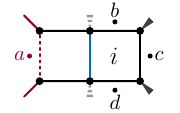}\hspace{-5pt}}{+}\fwboxR{70pt}{\fig{-2.25pt}{1}{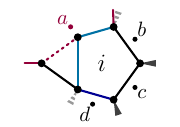}\hspace{-7.5pt}}{+}\fwboxR{70pt}{\fig{-2.25pt}{1}{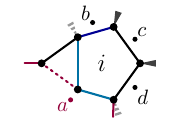}\hspace{-7.5pt}}\hspace{-5pt}\left.\rule{0pt}{26pt}\right\}\!\!\equivL\hspace{-5pt}\fwbox{65pt}{\fig{-2.25pt}{1}{4_ls_div_1}}\hspace{-5pt}\mathcal{D}^{i,\hspace{4pt}\r{a}}_{\hspace{-2pt}\r{a},b,c,d}.\hspace{-110pt}
\vspace{-6pt}\label{two_loop_d_int_defined}}

Actually, some care is required in the definition of these terms, as (\ref{two_loop_d_int_defined}) implicitly allows for cases involving a massless triangle integral,
\vspace{-6pt}\eq{\fig{-2.25pt}{1}{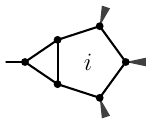},\vspace{-6pt}}
for which maximal cuts (and therefore master integrands) are ill-defined. For these cases, we may \emph{define} the integrands appearing in (\ref{two_loop_d_int_defined}) to be given by 
\vspace{-5pt}\eq{\fwbox{65pt}{\fig{-2.25pt}{1}{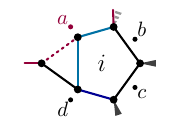}}\equivR\fwboxR{62pt}{\fig{-2.25pt}{1}{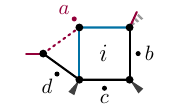}\hspace{-8pt}}\qquad\text{and}\qquad\fwbox{65pt}{\fig{-2.25pt}{1}{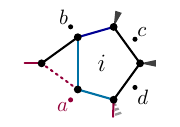}}\equivR\fwboxR{62pt}{\fig{-2.25pt}{1}{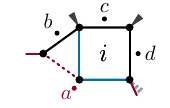}\hspace{-8pt}.}\label{boundary_conditions_for_d_ints}\vspace{-5pt}}

Combining the terms (\ref{two_loop_d_int_defined_1}) and (\ref{two_loop_d_int_defined}), we may write
\vspace{-6pt}\eq{\mathcal{A}^{(2)}_{\text{$\log^2$\hspace{-1pt}-div}}\equivR\sum_{\substack{a,b,c,d\\i}}\mathfrak{a}^{i}_{a,b,c,d}\Big[\Big(\sum_{\hspace{-2pt}\r{e}\notin\{a,b,c,d\}\hspace{-19pt}}\mathcal{I}^{i}_{a,b,c,d}\mathcal{I}^{\r{e}}_{\text{div}}\Big){+}\Big(\sum_{\hspace{-2pt}\r{e}\in\{a,b,c,d\}\hspace{-19pt}}\mathcal{D}^{i,\hspace{4pt}\r{e}}_{\hspace{-2pt}{a},b,c,d}\Big)\Big]\,.\vspace{-6pt}\label{a2_d1_v0}}
It will be useful to make this representation slightly more uniform by introducing 
\vspace{-6pt}\eq{\hspace{-0pt}\mathcal{F}^{i,\hspace{4pt}\r{a}}_{\hspace{-2pt}\r{a},b,c,d}\equivR\mathcal{D}^{i,\hspace{4pt}\r{a}}_{\hspace{-2pt}\r{a},b,c,d}{-}\big(\mathcal{I}^{i}_{a,b,c,d}\!\times\!\mathcal{I}^{{\r{a}}}_{\text{div}}\big)\vspace{-6pt}\label{finite_combo_defined}}
so that we may re-write (\ref{a2_d1_v0}) as
\eq{\begin{split}\mathcal{A}^{(2)}_{\text{$\log^2$\hspace{-1pt}-div}}&\equivR\sum_{\substack{a,b,c,d\\i}}\mathfrak{a}^{i}_{a,b,c,d}\Big[\Big(\sum_{\hspace{-2pt}\r{e}\notin\{a,b,c,d\}\hspace{-19pt}}\mathcal{I}^{i}_{a,b,c,d}\mathcal{I}^{\r{e}}_{\text{div}}\Big){+}\Big(\sum_{\hspace{-2pt}\r{e}\in\{a,b,c,d\}\hspace{-19pt}}\mathcal{D}^{i,\hspace{4pt}\r{e}}_{\hspace{-2pt}{a},b,c,d}\Big)\Big]\\
&=\sum_{\substack{a,b,c,d\\i}}\mathfrak{a}^{i}_{a,b,c,d}\Big[\mathcal{I}^{i}_{a,b,c,d}\mathcal{I}^{(1)}_{\text{div}}{+}\Big(\sum_{\hspace{-2pt}\r{e}\in\{a,b,c,d\}\hspace{-19pt}}\mathcal{F}^{i,\hspace{4pt}\r{e}}_{\hspace{-2pt}{a},b,c,d}\Big)\Big]\\
&\equivL\mathcal{A}^{(1)}_{\text{finite}}\,\mathcal{I}^{(1)}_{\text{div}}{+}\sum_{\substack{a,b,c,d\\i}}\mathfrak{a}^{i}_{a,b,c,d}\mathcal{F}^{i}_{\hspace{-2pt}{a},b,c,d}\,.\end{split}\label{a2_d1_for_ratio}}%

Finally, the $\log^4$-leading singularities are all equivalent to the tree amplitude, $\mathcal{A}^{(0)}$, and so we have
\vspace{-6pt}\eq{\mathcal{A}^{(2)}_{\text{$\log^4$\hspace{-1pt}-div}}\equivR\mathcal{A}^{(0)}\mathcal{I}^{(2)}_{\text{div}}\vspace{-6pt}\label{a2_d2}}
where the process-independent combination of integrands $\mathcal{I}^{(2)}_{\text{div}}$ are defined as
\vspace{-10pt}\eq{\begin{split}\mathcal{I}^{(2)}_{\text{div}}\equivR&\phantom{{+}}\Bigger{\sum}\hspace{-8pt}\fig{-2.25pt}{1}{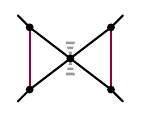}\hspace{-8pt}{+}\Bigger{\sum}\hspace{-12pt}\fig{-2.25pt}{1}{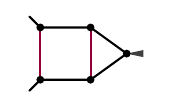}\hspace{-10pt}{+}\Bigger{\sum}\hspace{-12pt}\fig{-2.25pt}{1}{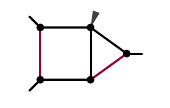}\hspace{-10pt}{+}\Bigger{\sum}\hspace{-12pt}\fig{-2.25pt}{1}{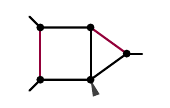}\hspace{-10pt}\\[-12pt]
&{+}\Bigger{\sum}\hspace{-6pt}\fig{-2.25pt}{1}{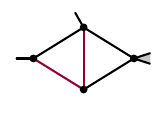}\hspace{-6pt}{+}\Bigger{\sum}\hspace{-6pt}\fig{-2.25pt}{1}{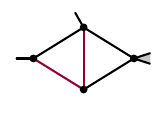}\hspace{-6pt}{+}\Bigger{\sum}\hspace{-6pt}\fig{-2.25pt}{1}{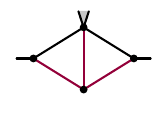}.\end{split}}

Putting everything together, we obtain an integrand-level representation of amplitudes in the basis of masters of \cite{Bourjaily:2026icc} given by 
\vspace{-6pt}\eq{\begin{split}\mathcal{A}^{(2)}&=\mathcal{A}^{(2)}_{\text{finite}}{+}\mathcal{A}^{(2)}_{\text{$\log^2$\hspace{-1pt}-div}}{+}\mathcal{A}^{(2)}_{\text{$\log^4$\hspace{-1pt}-div}}\\
&=\Big(\mathcal{A}^{(2)}_{\text{finite}}{+}\sum_{\substack{a,b,c,d\\i}}\mathfrak{a}^{i}_{a,b,c,d}\mathcal{F}^{i}_{\hspace{-2pt}{a},b,c,d}\Big){+}\mathcal{A}^{(1)}_{\text{finite}}\mathcal{I}^{(1)}_{\text{div}}{+}\mathcal{A}^{(0)}\mathcal{I}^{(2)}_{\text{div}}\,.\\[-10pt]\end{split}\label{a2_new}}
As we will see in the next section, a remarkable consequence of this representation of amplitudes will be to render the perturbative ratios of amplitudes \emph{manifestly} finite---computable without reference to any form of regularization or scheme dependence.

\newpage

\section{Manifesting Local Finiteness of Infrared-Safe Quantities}\label{sec_ratio_functions}

Consider the ratio of two perturbative scattering amplitudes, both involving the same number of external particles. We may define their perturbative \emph{ratios} by expanding this ratio order by order in $\hbar$ as in 
\vspace{-5pt}\eq{\frac{\mathcal{A}}{\mathcal{B}}\equivR\frac{\mathcal{A}^{(0)}{+}\hbar\,\mathcal{A}^{(1)}{+}\hbar^2\mathcal{A}^{(2)}{+}\ldots}{\mathcal{B}^{(0)}{+}\hbar\,\mathcal{B}^{(1)}{+}\hbar^2\mathcal{B}^{(2)}{+}\ldots}\equivL\,\mathcal{R}^{(0)}{+}\hbar\,\mathcal{R}^{(1)}{+}\hbar^2\mathcal{R}^{(2)}{+}\ldots\,.\vspace{-5pt}}
By multiplying both sides by $\mathcal{B}^{(0)}$ and matching orders, we can immediately see that\\[-8pt]
\vspace{-5pt}\eq{\mathcal{A}^{(L)}=\mathcal{R}^{(L)}\mathcal{B}^{(0)}{+}\mathcal{R}^{(L{-}1)}\mathcal{B}^{(1)}{+}\ldots{+}\mathcal{R}^{(0)}\mathcal{B}^{(L)}\,,\vspace{-5pt}}
so that 
\eq{\mathcal{R}^{(L)}\mathcal{B}^{(0)}=\mathcal{A}^{(L)}{-}\mathcal{R}^{(0)}\mathcal{B}^{(L)}{-}\ldots{-}\mathcal{R}^{(L{-}1)}\mathcal{B}^{(1)}\,.}
Introducing $\hat{\mathcal{B}}^{(L)}\equivR\mathcal{B}^{(L)}/\mathcal{B}^{(0)}$, we see in particular that
\eq{\begin{split}
\mathcal{R}^{(0)}\mathcal{B}^{(0)}&=\mathcal{A}^{(0)}\;\\
\mathcal{R}^{(1)}\mathcal{B}^{(0)}&=\mathcal{A}^{(1)}{-}\mathcal{R}^{(0)}\mathcal{B}^{(1)}\\
&=\mathcal{A}^{(1)}{-}\mathcal{A}^{(0)}\hat{\mathcal{B}}^{(1)}\\
\mathcal{R}^{(2)}\mathcal{B}^{(0)}&=\mathcal{A}^{(2)}{-}\mathcal{R}^{(0)}\mathcal{B}^{(2)}{-}\mathcal{R}^{(1)}\mathcal{B}^{(1)}\\
&=\mathcal{A}^{(2)}{-}\mathcal{A}^{(0)}\hat{\mathcal{B}}^{(2)}{-}\big(\mathcal{A}^{(1)}{-}\mathcal{A}^{(0)}\hat{\mathcal{B}}^{(1)}\big)\hat{\mathcal{B}}^{(1)}\\
&=\mathcal{A}^{(2)}{-}\mathcal{A}^{(0)}\hat{\mathcal{B}}^{(2)}{-}\mathcal{A}^{(1)}\hat{\mathcal{B}}^{(1)}{+}\mathcal{A}^{(0)}\hat{\mathcal{B}}^{(1)}\hat{\mathcal{B}}^{(1)}\\[-12pt]
\end{split}\label{one_and_two_loop_ratio_functions_spelled_out}}
and so-on. 

In the physics vernacular, \emph{the ratio function}\textsuperscript{\textregistered} is the name used to describe the case above where $\mathcal{A}$ is a \emph{ordered} N${}^k$MHV partial amplitude and $\mathcal{B}$ is the MHV partial amplitude with the same ordering, and both are evaluated for identical external momenta. This is convenient, as the MHV (super-)amplitude is in fact the Jacobian of the change of variables from spinors to (super) momentum-twistors, so that the MHV super-amplitude at tree-level is simply `$1$' in these variables \cite{ArkaniHamed:2009vw}; in particular, this means that $\mathcal{B}^{(0)}{=}1$ and $\hat{\mathcal{B}}^{(L)}{=}\mathcal{B}^{(0)}$ in the formulae above. Despite this convenience, however, we choose here to follow a broader discussion, allowing ratios of perturbative amplitudes involving the same N${}^k$MHV-degree, or different colour-orderings, or even perhaps considered for distinct sets of external momenta.\\[-6pt]

Provided all infrared (and/or ultraviolet) divergences may be organized according to
\vspace{-5pt}\eq{\mathcal{A}^{(L)}\equivL\mathcal{A}^{(L)}_{\text{finite}}{+}\sum_{q{=}1}^{L}\mathcal{A}_{\text{finite}}^{(L{-}q)}\mathcal{D}^{(q)}_{\text{div}},\vspace{-5pt}}
for \emph{any} form of the divergences gathered by $\mathcal{D}^{(q)}_{\text{div}}$, then it is not hard to show that $\mathcal{R}^{(L)}$ will be finite at every order of perturbation theory. The only crucial ingredient in this argument is that the divergences be \emph{universal} to both $\mathcal{A}$ and $\mathcal{B}$. (This criterion is substantially weaker than what is actually true for quantum field theories of massless particles; for example, this universality by itself does \emph{not} ensure that the logarithms of amplitudes will have bounded degrees of divergence---which follows from the specific ways in which $\mathcal{D}^{(L)}$'s actually arise from exponentiation of lower loop-orders (see \emph{e.g.}~\cite{Catani:1998bh,Catani:1999ss,Bern:2005iz}).)

If interpreted at the level of integrated, (necessarily) regularized amplitudes, the fact that ratio functions are finite is fairly trivial and arguably vacuous: obviously, in any renormalizable quantum field theory the divergences of amplitudes (viewed as coefficients of powers of $1/\epsilon$ in dimensional regularization, say) must be proportional to terms generated by lower loops, and so ratios between different helicity amplitudes, or even differently coloured amplitudes (for the same set of external momenta, at least) will be infrared (and ultraviolet) finite.\\[-6pt]

The finiteness of the ratio function appears much less trivial, however, when it can be realized at the level of loop integrands. Consider for example the completely prescriptive expansion of amplitudes at one loop in (\ref{a1_chiral_box_expansion}). Here, we have a representation of amplitudes of the form 
\eq{\mathcal{A}^{(1)}=\mathcal{A}^{(1)}_{\text{finite}}{+}\mathcal{A}^{(1)}_{\text{div}}=\mathcal{A}^{(1)}_{\text{finite}}{+}\mathcal{A}^{(0)}\,\mathcal{I}^{(1)}_{\text{div}}\,,}
where the finite parts involves integrands which are strictly, \emph{locally finite} \cite{Bourjaily:2021ewt} in four dimensions, and the divergences are identically prescribed for all amplitudes. Using this representation of amplitudes at one loop, ratio functions at one loop (\ref{one_and_two_loop_ratio_functions_spelled_out}) become---at the \emph{integrand-level}:
\vspace{-4pt}\eq{\begin{split}\mathcal{R}^{(1)}\mathcal{B}^{(0)}&=\mathcal{A}^{(1)}{-}\mathcal{A}^{(0)}\hat{\mathcal{B}}^{(1)}\\
&=\big(\mathcal{A}^{(1)}_{\text{finite}}{+}\mathcal{A}^{(0)}\mathcal{I}_{\text{div}}^{(1)}\big){-}\mathcal{A}^{(0)}\big(\hat{\mathcal{B}}^{(1)}_{\text{finite}}{+}\mathcal{I}^{(1)}_{\text{div}}\big)\\
&=\mathcal{A}^{(1)}_{\text{finite}}{-}\mathcal{A}^{(0)}\hat{\mathcal{B}}^{(1)}_{\text{finite}}\,.\\[-4pt]\end{split}\label{manifestly_finite_r1}}
This representation is \emph{manifestly} finite term-by-term, and requires no regularization of (or even reference to) the infrared divergent integrals appearing in individual amplitudes. As such, it is obvious that ratio functions must be entirely regulator (and hence also scheme) independent objects.  

Before moving on, it is worthwhile to mention that there exists another, quite distinct representation of ratio functions at one loop that manifests both their finiteness \emph{and} their dual-conformality. Specifically, this is provided by the `chiral octagon' basis of master integrals described in \cite{ArkaniHamed:2010gh}; this master integral basis cleanly separates IR-finite from IR-divergent master integrals, with the property that no linear combination of IR-divergent integrals is finite. In this basis, any IR-finite quantity (including, but not limited to the ratio function) must necessarily be spanned by the \emph{finite} subset of masters. 

Although this basis was used to write some ratio functions at one loop in closed form in \cite{ArkaniHamed:2010gh}, these integrals suffer from the fact that, as the masters individually involve more than four propagators, any cut supported in field theory will be supported by a large array of distinct master integrals. In particular, this means that the masters are \emph{far} from prescriptive: the integrals of the masters against the cuts supported by field theory is generically \emph{dense} for any amplitude. This makes it rather difficult to write amplitudes or ratio functions in terms of these integrals in closed form (without considerable cleverness of organization).

\subsection{Local Finiteness of Two-Loop Ratio Functions in sYM}\label{sec_local_finiteness_of_r2}

Until now, there has not been any representation of amplitudes at two or more loops which makes \emph{manifest} the finiteness of ratio functions as seen at one loop. The manifestly conformal representation of (\ref{a2_conformal}) \cite{Bourjaily:2015jna} reviewed in section~\ref{review_of_conformal_a2} does an admirable job of cancelling many of the divergences that appear in ratio functions (see \emph{e.g.}~\cite{Bourjaily:2018aeq,Bourjaily:2019jrk,Bourjaily:2019vby}), but falls short of actually resulting in representation that is infrared-finite term-by-term. Indeed, the computation of the six-point NMHV ratio function in \cite{Bourjaily:2019jrk} via explicit integration of each term demonstrates both how close and how far the representation of \cite{Bourjaily:2015jna} is from achieving this goal. 

The finiteness of ratio functions at two loops in fact becomes manifest in the completely prescriptive representation of (\ref{a2_new}). To see this, we simply use the form of amplitude integrands of (\ref{a2_new}) in the expansion for the ratio function given in (\ref{one_and_two_loop_ratio_functions_spelled_out}):
\eq{\begin{split}
\mathcal{R}^{(2)}\mathcal{B}^{(0)}=&\phantom{-}\mathcal{A}^{(2)}{-}\mathcal{A}^{(0)}\hat{\mathcal{B}}^{(2)}{-}\mathcal{A}^{(1)}\hat{\mathcal{B}}^{(1)}\\
=&\phantom{-}\big(\mathcal{A}^{(2)}_{\text{finite}}{+}\mathcal{A}^{(2)}_{\text{$\log^2$-div}}{+}\mathcal{A}^{(0)}\mathcal{I}^{(2)}_{\text{$\log^4$-div}}\big){-}\mathcal{A}^{(0)}\big(\hat{\mathcal{B}}^{(2)}_{\text{finite}}{+}\hat{\mathcal{B}}^{(2)}_{\text{$\log^2$-div}}{+}\mathcal{I}^{(2)}_{\text{$\log^4$-div}}\big)\hspace{-30pt}\\
&{-}\big(\mathcal{A}^{(1)}_{\text{finite}}{-}\mathcal{A}^{(0)}\hat{\mathcal{B}}^{(1)}_{\text{finite}}\big)\big(\hat{\mathcal{B}}^{(1)}_{\text{finite}}{+}\mathcal{I}^{(1)}_{\text{div}}\big)\\
=&\phantom{-}\big(\mathcal{A}^{(2)}_{\text{finite}}{-}\mathcal{A}^{(0)}\mathcal{B}^{(2)}_{\text{finite}}\big){-}\big(\mathcal{A}^{(1)}_{\text{finite}}{-}\mathcal{A}^{(0)}\hat{\mathcal{B}}^{(1)}_{\text{finite}}\big)\hat{\mathcal{B}}^{(1)}_{\text{finite}}\\
&{+}\big(\mathcal{A}^{(2)}_{\text{$\log^2$-div}}{-}\mathcal{A}^{(1)}_{\text{finite}}\mathcal{I}^{(1)}_{\text{div}}\big){-}\mathcal{A}^{(0)}\big(\hat{\mathcal{B}}^{(2)}_{\text{$\log^2$-div}}{-}\hat{\mathcal{B}}^{(1)}_{\text{finite}}\mathcal{I}^{(1)}_{\text{div}}\big)\\
=&\phantom{-}\mathcal{A}^{(2)}_{\text{finite}}{-}\mathcal{A}^{(0)}\mathcal{B}^{(2)}_{\text{finite}}{-}\mathcal{A}^{(1)}_{\text{finite}}\hat{\mathcal{B}}^{(1)}_{\text{finite}}{+}\mathcal{A}^{(0)}\hat{\mathcal{B}}^{(1)}_{\text{finite}}\hat{\mathcal{B}}^{(1)}_{\text{finite}}\\
&{-}\sum_{\substack{a,b,c,d\\i}}\mathfrak{a}^{i}_{a,b,c,d}\mathcal{F}^{i}_{\hspace{-2pt}{a},b,c,d}\;\;{+}\sum_{\substack{a,b,c,d\\i}}\hat{\mathfrak{b}}^{i}_{a,b,c,d}\mathcal{F}^{i}_{\hspace{-2pt}{a},b,c,d}\,.\\[-8pt]\end{split}\label{manifestly_finite_r2}}
To see that this representation involves only \emph{manifestly}-finite integrals term-by-term, the only non-trivial question involves the terms in the last line (\ref{manifestly_finite_r2}). 

Recall that $\mathcal{F}^{i}_{\hspace{-2pt}a,b,c,d}\equivR\sum_{\r{e}\in\{a,b,c,d\}}\mathcal{F}^{i,\hspace{4pt}\r{e}}_{\hspace{-2pt}a,b,c,d}$, where these integrands were defined by (\ref{finite_combo_defined}) to be the combinations
\vspace{-6pt}\eq{\hspace{-80pt}\mathcal{F}^{i,\hspace{4pt}\r{a}}_{\hspace{-2pt}\r{a},b,c,d}\equivR\hspace{-0pt}\fwboxR{75pt}{\fig{-2.25pt}{1}{d_abcd_1}\hspace{-5pt}}{+}\fwboxR{70pt}{\fig{-2.25pt}{1}{d_abcd_2}\hspace{-7.5pt}}{+}\fwboxR{70pt}{\fig{-2.25pt}{1}{d_abcd_3}\hspace{-7.5pt}}-\fwboxR{95pt}{\fig{-2.25pt}{1}{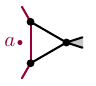}\hspace{-6pt}\times\hspace{-5pt}\fig{-2.25pt}{1}{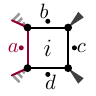}\hspace{-0.5pt}.\!}\hspace{-5pt}\hspace{-50pt}\vspace{-6pt}\label{finite_combination_spelled_out}}
This combination obviously cancels the soft-collinear divergence associated with $\ell\!\mapsto\!\r{p_a}$; what is less obvious is that this combination also cancels any soft(-but not collinear) divergence involving either massless leg adjacent to the edge with momentum $\r{p_a}$. Although we have not found any simple reason why this must be the case, it is straight-forward to verify that this is the case using the explicit forms of integrands defined in \cite{Bourjaily:2026icc}. (In the ancillary files for this work, we illustrate how this can be checked explicitly using randomly-chosen kinematics.) Because these combinations of integrands vanish in both collinear regions (everywhere as a function of the second loop momentum), they are \emph{locally finite} \cite{Bourjaily:2021ewt}, and therefore may be integrated without any use of regularization.\\[-6pt]

Although the basis of master integrals described in \cite{Bourjaily:2026icc} itself cleanly separates those with infrared divergences from those that are locally finite, it may at first seem surprising that \emph{new} locally finite integrands such as $\mathcal{F}^{i,\hspace{4pt}\r{a}}_{\hspace{-2pt}\r{a},b,c,d}$ (\ref{finite_combination_spelled_out}) exist---and are moreover crucial in manifesting the finiteness of infrared-safe observables such as ratio functions. But this should not, in fact, be entirely surprising: master integral bases such as those of \cite{Bourjaily:2026icc} are defined \emph{in terms of Feynman-like integrals}: those built from the propagators arising from a (connected) Feynman \emph{graph}. These obviously suffice for the description of \emph{amplitudes}, representable as they are in terms of (connected) Feynman diagrams; but ratio functions---even those defined with respect to a single planar ordering of external legs---necessarily con-fuse local $L$-loop contributions with those built as products of lower-loop contributions.

A master integral basis defined to encapsulate both local integrals and disconnected products thereof would necessarily be much larger than one sufficient to span merely $L$-loop amplitude integrands. The fact that \emph{divergent} combinations of two-loop integrands may be combined with divergent products of one-loop integrands to produce strictly, locally finite new masters is very interesting, and largely unexplored. 

It is worth noting that the combinations of integrands appearing in (\ref{finite_combination_spelled_out}) cannot be combined into any single set of propagators arising from any single Feynman graph. These integrals each involve at most three of the propagators chosen from
\eq{\hspace{-10pt}\begin{array}{@{}l@{}c@{}c@{}c@{}c@{}}
&\fig{-2.25pt}{1}{d_abcd_1}&\fig{-2.25pt}{1}{d_abcd_2}&\fig{-2.25pt}{1}{d_abcd_3}&\fig{-2.25pt}{1}{3_int_labelled_2}\hspace{-6pt}\times\hspace{-5pt}\fig{-2.25pt}{1}{4_int_i_labelled}\\
\x{\t{\ell_2}}{\r{a}}&\text{---}&\checkmark&\checkmark&\checkmark\\
\x{\r{\ell_1}}{\r{a\text{{\footnotesize+}}\hspace{-1pt}1}}&\checkmark&\text{---}&\checkmark&\checkmark\\
\x{\r{\ell_1}}{\r{a\hspace{2pt}\text{-}\hspace{0.5pt}1}}&\checkmark&\checkmark&\text{---}&\checkmark\\
\x{\r{\ell_1}}{\t{\ell_2}}&\checkmark&\checkmark&\checkmark&\text{---}\\
\end{array}\hspace{-20pt}}
but there is no local graph that contains all four. Nevertheless, this poses no obstruction to the direct integration via Feynman parameterization, say, of the combined integrand viewed as a form involving the union of their propagators.

\newpage
\section{Implementation of these Results in Mathematica}\label{sec_ancillary_files}

Both to sharpen the definitions of terms involved in the representation of amplitudes described here and also to provide some concrete tools for the manipulation of these expressions, we have prepared their implementation in \textsc{Mathematica} as ancillary files for this work as submitted to the \texttt{arXiv}. 

Although this implementation of course builds upon many of the tools made available in other works (see \emph{e.g.}~\cite{Bourjaily:2010wh,Bourjaily:2012gy,Bourjaily:2026adf,Bourjaily:2023uln}), the representation of amplitude integrands provided by this package is quite novel, based primarily upon \emph{seed} contributions to terms. These are briefly outlined below, and more thoroughly demonstrated in the accompanying walkthrough notebook.

\subsection{Primary Functionality of the Ancillary Package Toolbox}\label{sec_primary_functions}

The primary functions made available by the package are

\begin{itemize}
\item\fun{amplitudeIntegrandSeeds}\brace{\var{$n$}\pattern,\var{$k$}\pattern,\var{$\ell$}\pattern}: returns a \built{List} of three \built{List}s of \emph{seeds} representing the three groups of contributions to the \var{$n$}-point \var{$\ell$}-loop N${}^{\var{k}}$MHV amplitude integrand as given in (\ref{a1_chiral_box_expansion}) or (\ref{a2_new})---that is, the \built{First} \built{List} encodes the `finite' contributions, and so-on;
\item\fun{ratioFunctionSeeds}\brace{\var{$n$}\pattern,\var{$k$}\pattern,\var{$\ell$}\pattern}: returns a \built{List} of six \built{List}s of \emph{seeds} representing the six groups of terms appearing in the locally finite, prescriptive representation of the \var{$n$}-point \var{$\ell$}-loop N${}^{\var{k}}$MHV ratio function integrand as given in (\ref{manifestly_finite_r1}) or (\ref{manifestly_finite_r2});
\item\fun{conformalAmplitudeIntegrandSeeds}\brace{\var{$n$}\pattern,\var{$k$}\pattern,\var{$\ell$}\pattern}: returns a \built{List} of three \built{List}s of \emph{seeds} representing terms appearing in the conformal representation of two loop amplitude integrands according to (\ref{a2_conformal});
\item\fun{conformalRatioFunctionSeeds}\brace{\var{$n$}\pattern,\var{$k$}\pattern,\var{$\ell$}\pattern}: returns a \built{List} of terms encoding \emph{seeds} for the representation of ratio functions using the representation of amplitudes of (\ref{a2_conformal}), combined with a merge operation as described in \cite{Bourjaily:2015jna}.  
\item\fun{seedCoefficient}\brace{\var{seedTerm}\pattern}: returns the on-shell functions (represented by momentum-twistor \fun{superFunction} objects) which encode the \emph{seed} contribution's coefficients;
\item\fun{seedIntegrand}\brace{\var{seedTerm}\pattern}: returns an analytic expression for the master integrands in terms of momentum-twistor variables of the corresponding seed's integral, without symmetrization over $\ell_1\!\leftrightarrow\!\ell_2$; 
\item\fun{completeContributions}\brace{\var{seedTerm}\pattern}: returns a \built{List} of all inequivalent (cyclic) images of the corresponding seed's coefficients and master integrands;
\item\fun{recursedAmpTerms}\brace{\var{$n$}\pattern,\var{$k$}\pattern,\var{$\ell$}\pattern}: \r{lorem}
\item\fun{draw}\brace{\var{seedTerm}\pattern}: provides a graphical representation of \var{seedTerm}; 
\item\fun{componentFunction}\brace{\var{component}\pattern}\brace{\var{expression}\pattern}: \r{lorem} 
\item \fun{evaluate}\brace{\var{expression}\pattern}: tools to convert analytic expressions to numeric evaluations for kinematics specified by (use generated, reference values of, or randomly generated) momentum twistors.\\[-20pt]~
\end{itemize}

\newpage

\section{Conclusions and Future Directions}\label{sec_conclusions}
\addtocontents{toc}{\protect~\\[-38.5pt]}

At one and two loops, we have seen that complete prescriptivity may be achieved using master integrands which span the complete space of maximal-weight integrals, and that, provided these masters are prescriptively chosen in such a way that every cut supported by field theory is individually matched, this ensures the finiteness of ratio functions at the integrand level. These statements, demonstrated in the case of maximally supersymmetric Yang-Mills theory in the planar limit, `sYM', likely apply as well manifest the finiteness of many IR/IV-safe observables of more general quantum field theories at one and two loops, but it remains to be seen if such a structure is possible to higher orders of perturbation theory.

One natural direction for future work, therefore, would be to explore the infrared structure of amplitudes at higher loops, and further explore the tension between manifesting this structure and that of (dual-)conformality---or good power-counting, more generally (see \emph{e.g.}~\cite{Bourjaily:2015jna,Bourjaily:2020qca}). 

At three loops, the authors of \cite{Bourjaily:2017wjl} showed that a prescriptive basis of conformal master integrals at three loops exists for all N${}^k$MHV amplitudes of sYM, generalizing the fairly compact, closed formula for all MHV amplitude integrands given in \cite{ArkaniHamed:2010gh}. Even the explicit form known for MHV leaves much to be desired, however, as individual terms involve a variety of impurities and lower-weight pollution; it would be worthwhile to seek a set of masters spanning all conformal, individually pure/canonical master integrals at three loops (and beyond), and use these to represent amplitudes; we strongly suspect these formulae will not render manifest the finiteness of ratio functions, but hopefully their inadequacy on this front will help illuminate the role played by complete prescriptivity in this regard. 

Although we have shown that complete prescriptivity suffices to ensure the finiteness of ratio functions at two loops, this construction severely obfuscates their conformality. It would also be worthwhile to seek some generalization of the chiral octagons described in \cite{ArkaniHamed:2010gh} at higher loops that would prove sufficient to manifest both finiteness and conformality simultaneously, even if at the (considerable?) cost of non-prescriptivity. 

Finally, the existence of a manifestly finite representation of the ratio function begs the question of how difficult these integrands may prove to integrate analytically. This is especially the case for the novel finite integrands such as $\mathcal{F}^i_{a,b,c,d}$ defined in (\ref{finite_combination_spelled_out}), a class of functions which remains largely unexplored. Many of integrals that appear in the conformal representation of amplitudes have been integrated to various degrees of analyticity (see \emph{e.g.}~\cite{Spiering:2024sea}), and it would be worthwhile to extend this work to the non-conformal integrals that appear in the basis \cite{Bourjaily:2026icc}. 

\vspace{\fill}
\subsection*{Acknowledgements}~\\[-28pt]

The author thanks Jaroslav Trnka, Shruti Paranjape, and Taro Brown for helpful discussions and early collaboration on this work. This work was supported in part by a grant from the US Department of Energy (No.\ DE-SC00019066) and the J.~Robert Oppenheimer Visiting Professor Endowment at the Institute for Advanced Study, Princeton. This work was performed in part at Aspen Center for Physics, which is supported by National Science Foundation grant PHY-2210452.

\newpage
%\bibliographystyle{physics}
%\bibliography{amplitude_refs}
%\addtocontents{toc}{\protect~\\[-8pt]}\addcontentsline{toc}{section}{References}\addtocontents{toc}{\protect~\\[-42pt]\mbox{\vspace{0pt}}\protect\par}
%\end{document}
%
\addtocontents{toc}{\protect~\\[-24pt]}\addcontentsline{toc}{section}{References}\addtocontents{toc}{\protect~\\[-42pt]\mbox{\vspace{0pt}}\protect\par}
\providecommand{\href}[2]{#2}\begingroup\raggedright\endgroup

\end{document}